\documentclass[aip,reprint]{revtex4-1}

\usepackage{graphicx}
\usepackage{ amssymb }
\usepackage{amsmath}
\usepackage{amsfonts}
\usepackage{algorithm}
\usepackage{algpseudocode}
\usepackage{subcaption}
\usepackage{mathtools }
\usepackage{float}
\usepackage{placeins}
\usepackage{bbm}
\usepackage{bm}
\usepackage[export]{adjustbox}
\DeclareMathOperator*{\argmin}{arg\,min}

\renewcommand{\vec}[1]{\bm{#1}}

\draft 
\begin{document}


\title{Data-Driven Performance Measures using Global Properties of Attractors for Black-Box Surrogate Models of Chaotic Systems} 



\author{L. Fumagalli}
\email[]{luci.fumagalli@tu-ilmenau.de}
\affiliation{Technische Universität Ilmenau, Institute of Physics}

\author{K. Lüdge}
\affiliation{Technische Universität Ilmenau, Institute of Physics}

\author{J. de Wiljes}
\affiliation{Technische Universität Ilmenau, Institute of Mathematics}
\affiliation{Lappeenranta-Lahti University of Technology, School of Engineering Science}
\author{H. Haario}
\affiliation{Lappeenranta-Lahti University of Technology, School of Engineering Science}

\author{L. Jaurigue}
\affiliation{Technische Universität Ilmenau, Institute of Physics}


\date{\today}

\begin{abstract}
\textbf{Abstract }

In climate systems, physiological models, optics, and many more, surrogate models are developed to reconstruct chaotic dynamical systems. We introduce four data-driven measures using global attractor properties to evaluate the quality of the reconstruction of a given surrogate time series. The measures are robust against the initial position of the chaotic system as they are based on empirical approximations of the correlation integral and the probability density function, both of which are global properties of the attractor. In contrast to previous methods, we do not need a manual fitting procedure, making the measures straightforward to evaluate. Using a hypothesis testing framework, we can systematically find and reject surrogate models whose reconstructions significantly differ from the true system. Further, we show that the measures can be used as a statistical ranking metric, which in practice allows for hyperparameter optimization. Applying our measures to reservoir computing with a low number of nodes, we demonstrate how the measures can be used to reject poor reconstructions and use them as a tool to find an optimal spectral radius for which considerably fewer solutions are rejected and the overall quality of the solution improves.
\end{abstract}


\maketitle 

\textbf{
Surrogate models are frequently used in climate systems, physiological models, optics, and other fields to reconstruct chaotic dynamics \cite{BAK96,ZHA08,DUR23}. For chaotic systems, small changes can lead to time series that appear similar in the short term but are uncorrelated over the long term (butterfly effect). Weather forecasts serve as an intuitive example — while short-term predictions are reliable, forecasting months ahead appears impossible as small differences 
in the simulation lead to seemingly random changes. In this paper, we want to assess the quality of the reconstruction of a surrogate model using only time series from the surrogate model (black box assumption). Taking the weather example, this means evaluating the quality of our physical description using only the generated weather forecast (time series reconstruction). Conventional machine learning techniques compare the reconstruction to the ground truth directly, which works for short-term but not for long-term time series due to their chaotic nature. We introduce four data-driven measures\footnotemark[2] for time series that capture the global behavior of the system, analog to analyzing climate trends rather than direct weather forecast. In contrast to previous methods, our measures do not need any manual intervention of choosing fitting ranges and/or embeddings\cite{ROS93, KAN03}. Using a statistical framework, we provide a strategy to systematically find and reject surrogate models whose reconstructions significantly differ from the true global behavior. Further, we show that the measures can be used to find an optimal optimization parameter for surrogate models. The measures have the potential to lead to new insights into the creation of accurate surrogate models \cite{HAL19,YAD25,JAU24a,KOE24,FLE25,GIL23,OEZ23,THO25}. For reservoir computing, a form of machine learning, we demonstrate how our measures can be used to reject poor reconstructions and use the measures as a tool to find an optimal reservoir optimization parameter for which considerably fewer solutions are rejected and the overall quality of the solution improves.
}

\section{Introduction}
In this paper, we address new measures\footnotemark[2] to determine the accuracy of a chaotic surrogate model. For this, we use global attractor properties such as the correlation integral and the probability density function. In particular, we focus on the use of these measures for classifying the performance of surrogate models which have been generated in a data-driven approach. 

\footnotetext[2]{We use the word ''measure'' to refer to performance metrics. We do not mean measure in a mathematically rigorous sense.}

Chaotic dynamics are ubiquitous in nature; examples are found in climate systems, physiological models, optics, and many more \cite{BAK96,ZHA08,DUR23}. 
In many examples, there is a desire to predict or control these dynamics.
When the underlying governing equations are not known, a common approach is to develop data-driven surrogate models using methods from statistical modeling and machine learning\cite{ZHA23k,DUR23}. To evaluate the accuracy of the surrogate models, there is a need for performance metrics that capture the global properties of the target system. Often, particularly when machine learning methods are used, the surrogate model is treated as a black box and one only has access to the resulting trajectories. The inherent sensitivity of chaotic dynamics to small variations in initial conditions means that these trajectories will vary from the reference trajectory on short to intermediate timescales, regardless of the quality of the surrogate model. This means that commonly used error measures, such as the valid time\cite{PAT18a} or the mean square error, are insufficient to judge if the global properties of the target system have been reproduced.


Measures that capture the global attractor properties include\cite{KAN03} the 
correlation dimension\cite{GRA83}, the Lyapunov exponents\cite{WOL85}, the Wasserstein distance\cite{MEM11} and the Kullback–Leibler divergence\cite{KUL51}. 
Although these are well-established measures their use is often hindered by factors such as the computational expense, the dependence on the choice of fitting ranges and/or embeddings\cite{ROS93, KAN03} as well as restrictions on the smoothness of the time series. 
Furthermore, the way in which such measures are typically used does not take into account the inherent variation of the reference system in a practical manner.

In this work, we introduce a statistical framework to four measures related to the correlation integral and the probability density function of chaotic trajectories. The first measure is the Generalized Correlation Integral, which was introduced in \cite{HAA15}. The second is the Total Variation, which is based on the probability density distribution of trajectories from the reference system and the surrogate model. The third measure, the Attractor Exclusion, we introduce here as a measure of how much the trajectories of the surrogate models lie outside of the space of the attractor of the reference system. The fourth measure is the Attractor Deviation, which is a measure of the difference in the space occupied by two attractors. For the Attractor Deviation we build on the definition introduced in \cite{JAU24a}. 
Throughout the paper, we assume that the reference trajectories are from chaotic systems that evolve close to an attractor. By definition, trajectories in a small environment around the attractor will remain close to the attractor. Thus, the trajectories, and hence the measures, reflect the long-term behavior of the chaotic reference system. 


In contrast to current approaches, our key contribution is that we apply a statistical framework to categorize surrogate models into rejected and accepted, analogous to hypothesis testing. We use the above-mentioned measures as our test statistics, for which we can define a threshold. Any surrogate model whose test statistic is above the threshold is deemed to be significantly different from the reference system. Importantly, due to our statistical approach and the measures being derived from the probability distributions of trajectories, the test time series need not be smooth. This means that our measures can be applied to temporally sparse reconstructions of dynamical systems.  Empirically, we find that the measures can be used to statistically rank solutions.

The evaluation of time series and the determination of the thresholds are data-driven. One exception is the General Correlation Integral, where an analytical expression for the threshold is given. To evaluate our measures and to create the thresholds, only access to time-discretized realizations of finite trajectories is needed and never any further information about the underlying model. This means that our approach is widely applicable, as well as being computationally tractable. Applying our statistical measures to studies where chaotic attractor reconstruction was performed has the potential to lead to new insights into the creation of accurate surrogate models \cite{HAL19,YAD25,JAU24a,KOE24,FLE25,GIL23,OEZ23,THO25}.

The measures introduced in this paper can be applied to time series independent of the manner in which they are generated. 
A particularly relevant use-case is surrogate models generated via data-driven approaches such as statistical modeling and machine learning.
We apply our measure to the latter, choosing a family of reservoir computing surrogates, where the underlying design is inspired by the possibility of using a physical systems for the nonlinear transformations.
 Creating surrogate models with reservoir computing has the potential to create energy-efficient and inexpensive real-world implementations \cite{JAU24a}. 

The paper is structured as follows. The models and numerical simulation parameters can be found in section \ref{sec:target_model}. In section \ref{sec:measures}, we introduce the statistical framework as well as the measures. We validate that the measures are able to distinguish a trend in model performance in section \ref{sec:validation}. Finally, we briefly introduce reservoir computing (section \ref{sec:reservoir_computing}) and apply our methods to reservoir computing (section \ref{sec:results}).

\section{Methods}

\subsection{Reference Model}\label{sec:target_model}

Throughout this paper, we use the Lorenz 63 system, which is a simplified model for atmospheric convection \cite{LOR63}, as the reference system for demonstrating our measures and the statistical framework. However, our approach is generally applicable to chaotic systems. In the supplemental material, we show results for four additional low-dimensional chaotic systems, namely Sprott B, Sprott C, Sprott D\cite{SPR94}, and Roessler76\cite{ROE76}.

The Lorenz 63 system is described by a set of ordinary differential equations, 
\begin{align}
\frac{du_x}{dt} &= \theta_x^\text{ref} \times (u_y - u_x) \\ 
\frac{du_y}{dt} &= u_x \times (\theta_y^\text{ref} - u_z) - u_y \\
\frac{du_z}{dt} &= u_x \times u_y - \theta_z^\text{ref} \times u_z,
\end{align}
which can exhibit chaotic dynamics. For the reference model, we use the common choice of parameters, $\vec \theta^\text{ref} = (10\;28\;\frac{8}{3})^T$, which gives rise to chaotic dynamics on a double-lobed attractor. The time series data needed for defining and testing the performance measures is generated by simulating the Lorenz equations using an explicit Runge Kutta solver of order 5(4) from scipy using Dormand-Prince pair formula\cite{DOR80} and adjustable step size to control the 4 order error with additional interpolation\cite{SHA86}, i.e. scipy.integrate.RK45\cite{scipy}. An equal distanced solution was simulated with a small integration time of $h = 1\times 10^{-3}$ to ensure a small integration error. We down-sample the simulated time series by a factor of 20 to generate sequences of $N_O = 5000$ observations with time-steps of $\Delta t = 20 \times h = 0.02$. This choice of $\Delta t$ was made such that the long-term behavior can be captured, while simultaneously keeping the number of data points relatively small and retaining the short timescale dynamics. However, the results are not contingent on this choice. In the supplemental, we show results for a more sparse sampling in time, $\Delta t_\text{Sparse} = 50 \times \Delta t$.


\begin{figure}[ht!]
    \centering
    \vspace{0.5cm}
    \begin{subfigure}{0.51\textwidth}
        \centering
        \includegraphics[width=0.55\textwidth]{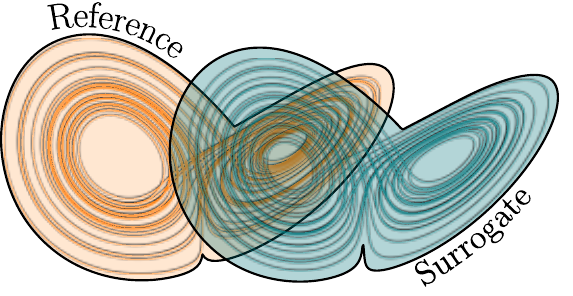}
        \caption{time series}
        \label{fig:definition_of_support}
    \end{subfigure}
    \begin{subfigure}{0.50\textwidth}
        \centering
        \includegraphics[width=0.43\linewidth]{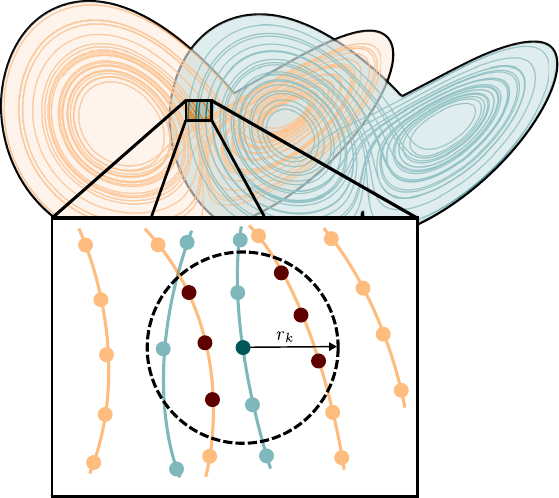}
        \caption{GCI}
        \label{fig:heikki}
    \end{subfigure}
    \begin{subfigure}{0.51\textwidth}
        \centering
        \includegraphics[width=0.5\textwidth]{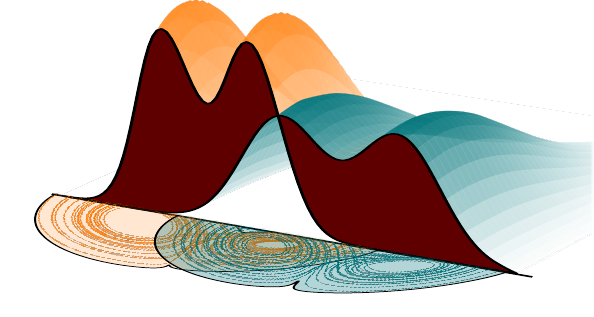}
        \caption{TVar}
        \label{fig:total_variation}
    \end{subfigure}
    \begin{subfigure}{0.51\textwidth}
        \centering
        \includegraphics[width=0.43\textwidth]{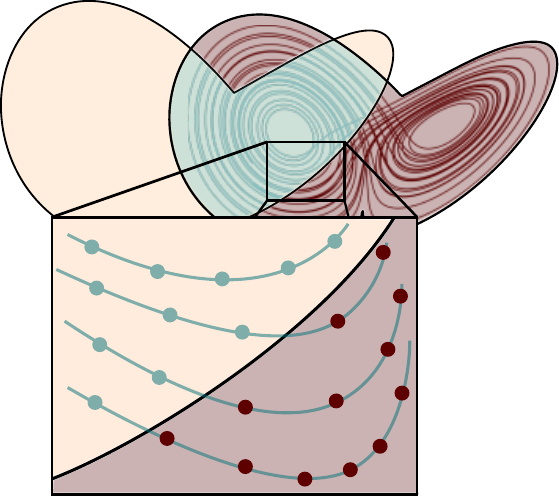}
        \caption{AExc}
        \label{fig:attractor_exclusion}
    \end{subfigure}
    \begin{subfigure}{0.51\textwidth}
        \centering
        \includegraphics[width=0.48\textwidth]{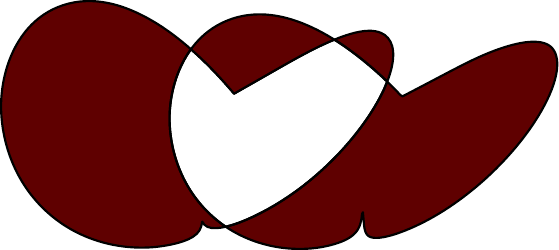}
        \caption{ADev}
        \label{fig:attractor_deviation}
    \end{subfigure}
    \caption{(\subref{fig:definition_of_support}) Input data of our measures. Orange and blue represent the reference and surrogate time series, respectively. The solid black outlines indicate the total attractor space. (\subref{fig:attractor_deviation} - \subref{fig:heikki}) Visualizations of the four measures. The dark red area is the quantity each measure is capturing. The ADev measures the difference of the attractors, the AExc counts all surrogate values outside the reference attractor, TVar uses the difference in probability density, and GCI captures the number of neighboring reference points for all data points from the surrogate trajectory.}
    \label{fig:overall}
\end{figure}

\subsection{Measures}\label{sec:measures}

The goal of this paper is to define computationally inexpensive measures to evaluate if a given time series has similar properties to, or can be considered as being a realization of the reference system. Due to the inherent variations that can arise in trajectories of the chaotic surrogate systems, we approach this goal statistically using techniques akin to hypothesis testing. 

In the following sections, we introduce four measures, which take the surrogate and reference time series and output positive real values. The surrogate and reference time series will be denoted as $U = \left(\vec u_t\right)_{t\in1,\ldots,N}$ and $U^\text{ref} = \left(\vec u_t^\text{ref}\right)_{t\in1,\ldots,N\text{ref}}$, where $\vec u_t$ and $\vec u_t^\text{ref}$ is the sample at time $t$ for the surrogate and reference trajectories, respectively.

We switch to a probabilistic framework, and thus, the measures are treated as random variables to model their uncertainty in the initial conditions of the reference system. To emphasize the randomness, we will refer to the measures as the test statistic. The distribution of the test statistic is determined either empirically or theoretically. Using this distribution,  we define a threshold such that 95\% of the reference time series fulfill this threshold. 
To evaluate a surrogate time series, we calculate the test statistic's value.
The surrogate time series is rejected if the test statistic's value exceeds the threshold. 
A time series that exceeds the threshold is considered significantly different from the reference system. 
By definition, only $100\% - 95\% = 5\%$ of reference time series would be rejected, which inspires high confidence in the above statement.
Identical to hypothesis testing, the reverse statement is not necessarily true, i.e. a time series from a different system can perform better using the test statistic than the reference system.

Apart from a reference time series, our measures do not need any knowledge of the underlying dynamics of the reference system (black box assumption).
However, the reference time series must be appropriately long to capture the distribution of the test statistic well. Since we assume that the reference time series is on the attractor and our measures capture the attractor properties, we can use multiple different time series from the reference system instead of a single one, i.e. we assume ergodicity on the attractor. The properties when ergodicity is fulfilled are known \cite{ECK85}. For the Lorenz 63 system, ergodicity has been proven\cite{TUC99}.

The first test statistic is the Generalized Correlation Integral (GCI), which we will use as a benchmark for the performance of the other test statistics as it is already established \cite{HAA15, SPR19}. For comparison with other works, we also define the valid prediction time, which is a measure of short-term forecasting performance.

\subsubsection{Generalized Correlation Integral (GCI)}

The first test statistic is the  Generalized Correlation Integral (GCI), which was introduced in \cite{HAA15} as validation for the first version of Generalized Correlation Integral Vector's (GCIV) distribution\footnote[3]{Newer versions include additional features such as the ID (Intrinsic Dimension, based on ratios of distances) or the Chamfer distance, which give rise to promising result by improve the measure's performance as well as reducing the number of data points needed to evaluate the measure reliably.}. 
In this paper, we will shift our focus and use the GCI directly as the main test statistic.

The GCI is calculated using the components of the GCIV, we therefore first give the definition of the GCIV. For a fixed set of radii $(r_i)_{i \in 1, \dots, R} \in \mathbb{R}^R$, number of observed samples $N^\text{O}$, and number of reference data samples $N^\text{Ref}$, the $k$-th components of the GCIV are defined as
\begin{align}\label{Eq:GCVI}
y_k(U, U^\text{ref}) = \frac{1}{N^\text{O}  N^\text{ref}} \sum_{i= 1}^{N^\text{O}} \sum_{j = 1}^{N^\text{ref}} \#(\|U_i - U^\text{ref}_j\|_2 < r_k),
\end{align}
where the function $\#(\text{statement})$ is 1 if the statement is true and 0 if not.
Throughout this paper, we choose $N^\text{ref} = 100 \times N^\text{O}$ and $N^\text{O}=5000$. 

The GCIV calculates the mean of how often the distance between any combination of points from the surrogate and reference time series is below the radius $r_k$. 
Figure \ref{fig:heikki} illustrates the inner summation. 
Orange and turquoise represent the reference and surrogate time series, respectively. 
A magnified view is provided below.
The continuous dynamics is represented by a line, while the observations are dots on the line.
The outer sum iterates over all turquoise points. Let the dark turquoise point in the center of the magnified view be the current selected point of the outer sum.
In the inner sum, we search for all orange points that are closer than $r_k$ away from the selected point, which we will mark as dark red.
The inner summation then counts the number of dark red points, while the outer repeats this process for all surrogate data points.

The GCIV can be used to differentiate the reference model from differently parametrized models \cite{HAA15} (see appendix \ref{App:GCI}). 
If the reference system and observed system are the same except for the initial condition, $r \rightarrow 0$, and $ N^\text{ref} = N^\text{O} \rightarrow \infty$, the GCIV is equivalent to the definition of the empirical correlation integral, and thus, closely related to the correlation dimension of the attractor \cite{HAA15, GRA83}. 

The GCI test statistic is defined as
\begin{align}
d_\text{GCI}(U, U^\text{ref}) = (\vec{\mu} - \vec y(U, U^\text{ref})) \bm \Sigma^{-1} (\vec{\mu} - \vec  y(U, U^\text{ref})),
\label{eq:GCI}
\end{align}
where $\bm \Sigma \in \mathbb{R}^{R \times R}$ is the covariance matrix and $\vec \mu \in \mathbb{R}^R$ the mean from the GCIV distribution, $\vec y$ \cite{HAA15}.
This test statistic is distributed according to a chi-squared distribution with $R$ degrees of freedom if both the surrogate and the reference system are from the same system and the data points are independent. The GCIV fits into the definition of a U-statistic, which guarantees $\vec y$ to converge towards a multivariate Normal distribution, supposing only weakly dependent data \cite{BOR01}.

For Eq.~(\ref{eq:GCI}) to be chi-square distributed, the samples must be (weakly) independent. This is not generally the case for samples on the trajectory of a dynamical system, for example, samples with a small temporal separation can be highly correlated. To ensure independent samples in a chaotic system, it suffices to decrease the sampling rate. We use a subsampling rate $c_\text{sub} = 1/15$ Thus, we subsample both the reference and surrogate time series by taking $N^\text{ref} \times c_\text{sub}=5000 \times 100 / 15 \approx 33333$ and $N^\text{O} \times c_\text{sub} = 5000 / 15 \approx 333$ random samples of the time series, respectively. 

To use the test statistic, we empirically characterize $\bm \Sigma$ and $\vec{\mu}$ for the reference system according to algorithm \ref{alg:heikki_ref}. We created for each the surrogate and reference system 300 time series. The $l$-th reference and $s$-th surrogate time series at the $t$-th time step are denoted by $\vec u _t^l$ and $\vec u _t^s$, respectively. To find appropriate radii, we calculate the Euclidean distance between the surrogate and reference time series for all time steps, $\bm D_{t_1, t_2}^{s, l} = \|\vec u_{t_1}^s - \vec u_{t_2}^l \|_2$.
The radii are then defined automatically by an exponential scale;
\begin{align}
    r_k = b \times (a/b)^{{k}/{(R+1)}}, k \in 1, \ldots, R
\end{align}

where the scale is between $\tilde a = a + \Delta$ and $\tilde b = b - \Delta$ with $\Delta = (b - a)/50$, $a =  \max_{s, l, s \neq l} \left( \min _{t_1, t_2} D_{t_1, t_2}^{s, l} \right) $, and $ b =  \min_{s, l, s \neq l} \left( \max_{t_1, t_2} D_{t_1, t_2}^{s, l} \right) $. The scale is chosen to be as large as possible without having too few samples in the tails. The exponential scale can be motivated by the connection to the correlation integral. 
Strictly speaking, in our setup the distance where $l = s$ could be included, we decided against it to be consistent with \cite{HAA15}. 

\begin{figure*}[htb!]
    \centering
    \includegraphics[width=0.9\linewidth]{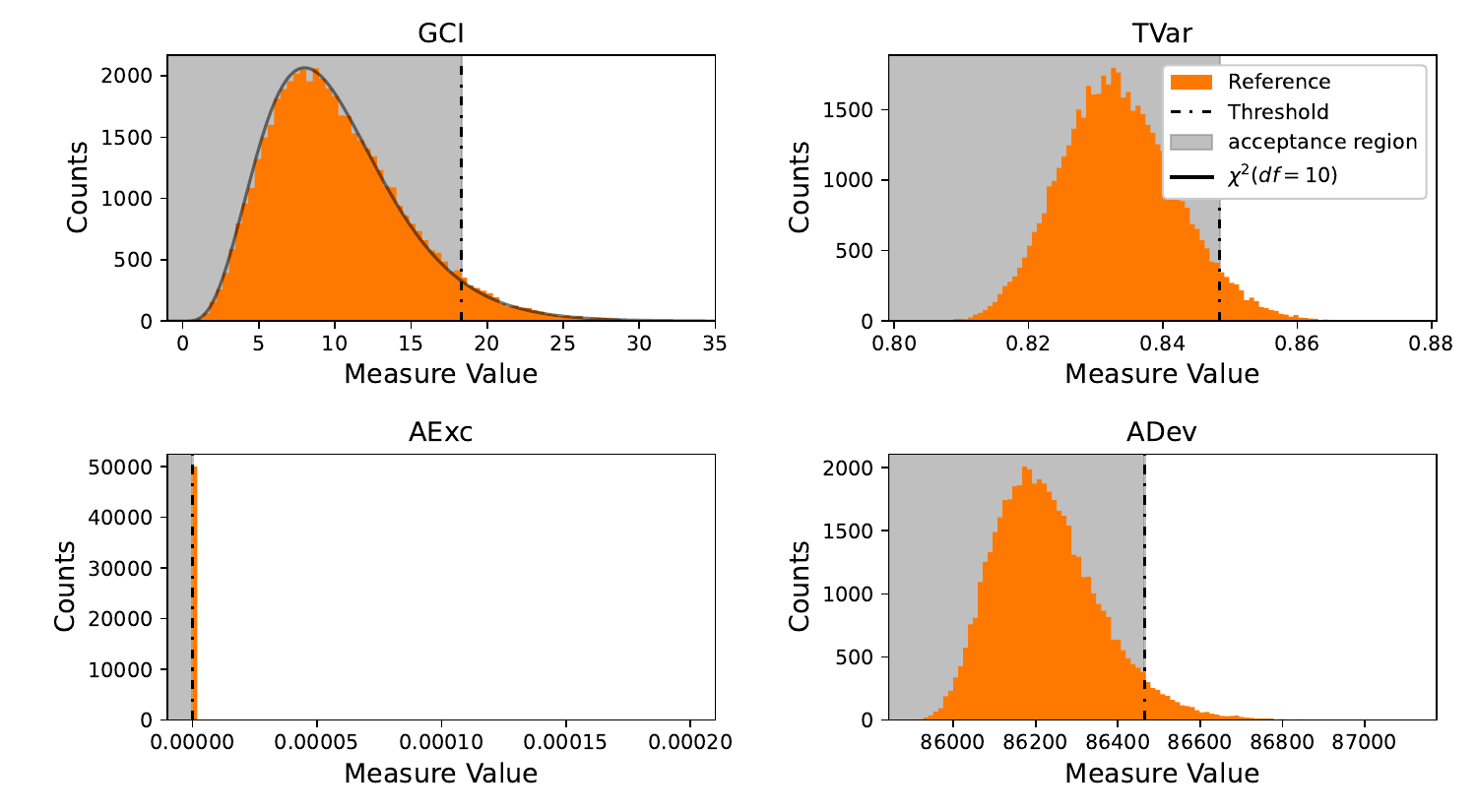}
    \caption{Histograms for the four measures when, in place of the surrogate time series, realizations of the reference Lorenz 63 are used. The threshold is shown as the dashed vertical line. It is determined by the 95 quantile, except for the GCI where it can be calculated since the GCI's distribution is governed by a $\chi^2$ distribution with $10$ degrees of freedom (solid black line). Any value that is larger than the threshold is rejected and deemed to be significantly different from the reference system, while every value below is accepted (acceptance region). Thus, for a perfect surrogate model, 95\% of its time series would be accepted. We used an observation time of $\Delta t = 0.02$, $50000$ random Lorenz 63 time series with $N_O=5000$ observations, and a single fixed realization of the reference series for the GCI. For all parameters see table \ref{tab:parameteres} upper row and for the threshold see table \ref{tab:thresholds}.}
    \label{fig:heikki_chi2}
\end{figure*}

To evaluate the test statistic, it is sufficient to use a single reference time series if it is sufficiently long, i.e., initial conditions of the reference time series have a small impact. Figure \ref{fig:heikki_chi2}(top left) shows Eq.~(\ref{eq:GCI}) evaluated for different time series from the reference model, with a single fixed reference time series to calculate the GCIV (Eq.~(\ref{Eq:GCVI})). The data fits the chi-square distribution from theory well, justifying our assumptions. 

Using the chi-squared distribution, a threshold can be defined analytically such that $95\%$ of test time series from the true system are below this threshold (see figure \ref{fig:heikki_chi2}). The threshold is defined as $\text{Threshold} = \text{icdf}_{\chi^2_{10}}(0.95) = 18.307$, where icdf is the inverse cumulative distribution function. Thus, for any time series that is above the threshold, we can reject it with a high probability of not being from the reference system. Although no data from the reference system is needed to calculate the threshold, it is possible to test for a good choice of the subsampling rate by comparing the empirical distribution to the probability density function of the $\chi^2$-distribution (see \ref{fig:heikki_chi2} solid black line). For this reason, the GCI needs much fewer data points to be used for reliable statistics than the following measures.

\subsubsection{Total Variation (TVar)}

The total variation measure is the empirical approximation of the total variation distance between the reference probability density distribution, $\pi(\vec u)$, and the surrogate system probability density distribution, $q(\vec u)$:
\begin{align}
    d_\text{TVar}(\pi, q) = \frac{1}{2}\int_{\mathbb{R}^D} \left| q(\vec u) - \pi(\vec u) \right| d \vec u.
\end{align}
Figure \ref{fig:total_variation} depicts the total variation as the dark red volume between the two probability density functions. It is zero if the probability density functions are the same. 

To approximate the above expression, the probability density distribution is approximated by histograms for both the surrogate and the reference system. 
To create the bins of the histogram, 60 time series with $201 \times 10^5$ time steps are simulated with $\Delta t = h = 0.001$ for the Lorenz system. To ensure that the time series starts at the attractor the first $1 \times 10^5$ time steps are ignored. We assume that all systems are ergodic on the attractor, and thus, we can use multiple time series to construct the histogram. Ergodicity on the attractor is proven for the Lorenz63 \cite{TUC99}. The simulation is used as an initial guess for the bins. The maximum and minimum along each dimension are determined. The space between the minimum and maximum is split into 234 equally spaced bins. Further, we expand the space by 16 bins in each direction to account for not covering the full space with our initial guess. To create the final histogram, $3600$ time series are generated of length  $201 \times 10^5$  where the first  $1 \times 10^5$ time steps are ignored. The same bins are used for the surrogate time series. Let $\vec \alpha(\vec u)$ be a multi-dimensional index to address all bins. Let $\mathcal{H}_{\vec \alpha}(\{\vec u_1, \vec u_2, \ldots, \vec u_N\})$ be the corresponding count of the histogram for the bin addressed by $\vec \alpha$. Then the Total Variation (TVar) test statistic is defined as

\begin{align}
    d_\text{TVar}\left(U, U^\text{ref}
\right) = \frac{1}{2} \sum_{\vec\alpha} \left| \mathcal{H}_{\vec \alpha}(U) - \mathcal{H}_{\vec \alpha}(U^\text{ref}) \right|.
\end{align}
To determine the threshold, we require the distribution of this test statistic for different realizations of the reference system. The distribution of 50000 simulated time series of our reference system is shown in  Fig.~\ref{fig:heikki_chi2}(top right). From this, we determine the 95 quantile empirically. 

\subsubsection{Attractor Exclusion (AExc)}

The Attractor Exclusion (AExc) test statistic measures how many samples are outside the attractor space of the reference system. For that, the attractor shape is approximated by creating a histogram for the reference system. We will use the same histogram from the previous section, $\mathcal{H}_{\vec \alpha}(U^\text{ref})$, where $\vec\alpha$ is a multi-index, to index the bins. Let us denote the bin of the sample $\vec u$ at time $t$ as $\vec \alpha(\vec u)$. The attractor exclusion is defined by
\begin{align}
    d_\text{AExc}\left(U, U^\text{ref}\right) = \frac{1}{N}\sum_{t=1}^N \#\left({\mathcal{H}_{\vec \alpha(U_t)}(U^\text{ref}) = 0}\right),
\end{align}
where the function $\#(\text{statement})$ is 1 if the statement is true and 0 if not. The histogram of the surrogate data uses the same bins as the histogram for the reference.
 The attractor exclusion counts what percentage of the samples are outside the definition space of the attractor. Figure \ref{fig:attractor_exclusion} gives a visualization of the attractor exclusion. The attractor space is represented by the black outlines. The red boxes represented a histogram of the surrogate time series. Summing up the boxes and dividing by the total number of samples results in the attractor exclusion. 

Similar to the TVar, we generate the distribution of the AExc test statistic for the reference system by simulating 50000 times series of the reference system. The resulting distribution is shown in Fig.~\ref{fig:heikki_chi2}(bottom left). The acceptance threshold is determined empirically from the 95 quantile of this distribution.

\subsubsection{Attractor Deviation (ADev)}

The Attractor Deviation (ADev) was introduced in \cite{JAU24a} and is similar to the definition used in \cite{ZHA23e}. It measures the deviation in attractor shape between the surrogate and reference system (see figure \ref{fig:attractor_deviation}). We use the same histogram from the previous sections, $\mathcal{H}_{\vec \alpha}(U^\text{ref})$, where $\vec \alpha$ is a multi-index, to index the bins. The ADev compares the support of both histograms. The support is the area, where the probability density is non-zero. The ADev is defined as
\begin{align}
    d_\text{ADev}\left(U, U^\text{ref}\right) = \frac{1}{2} \sum_{\vec\alpha} &\big|\#(\mathcal{H}_{\vec \alpha}(U) > 0) \nonumber \\ 
    &- \#(\mathcal{H}_{\vec \alpha}(U^\text{ref}) > 0) \big|,
\end{align}
where the function $\#(\text{statement})$ is 1 if the statement is true and 0 if not.

Again, the distribution of the ADev test statistic for the reference system is determined by simulating 50000 time series (see Fig.~\ref{fig:heikki_chi2}(bottom right)) and the threshold is given empirically by the 95 quantile. 

\subsubsection{Valid Prediction Time (VPT)}

In contrast to the previous measures, the valid prediction time quantifies the short-term predictive quality of the surrogate system in comparison to a particular reference trajectory. There are multiple slightly varying definitions of the VPT time or similar measures such as the forecast horizon \cite{HAL19} and the valid time \cite{PAT18a}. Here we use the definition from \cite{KOE23}: 
\begin{align}
    \delta_{\text{VPT}}\left(U, U^\text{ref}\right)  &= \argmin_t \Big[ {\lVert U _t - U_t^\mathrm{ref} \rVert^2} \nonumber\\
    &\big/\Big({\frac{1}{N} \sum_{i} \lVert U _i^\mathrm{ref} - \frac{1}{N} \sum_{j} U _j^\mathrm{ref} \rVert^2}\Big) > 0.4 \Big].
\end{align}
 The denominator is the sum of the empirical variance over all components. The valid prediction time is the first time step, where the surrogate time series is significantly different from the reference time series, where significance is defined via the variation of the time series. 

The VPT time depends strongly on the initial conditions of the reference time series and does not provide information about the long-term properties of the surrogate system. We include this measure in our study for comparison with previous works.

\subsection{Validation}\label{sec:validation}

In this section, we validate the measures introduced above by their ability to distinguish time series generated from miss parameterized systems and time series from the reference system that have been altered. For the generalized correlation integral vector, on which the GCI is built, the ability to distinguish miss parameterized systems was shown for a variety of dynamical reference systems in \cite{HAA15}. Here we focus on one dynamical reference system, the Lorenz 63 system, to highlight the comparative performance of our four measures. Results for other dynamical reference systems are given in the supplemental material.

%
\begin{figure}[h]
    \centering
    \includegraphics[width=\linewidth]{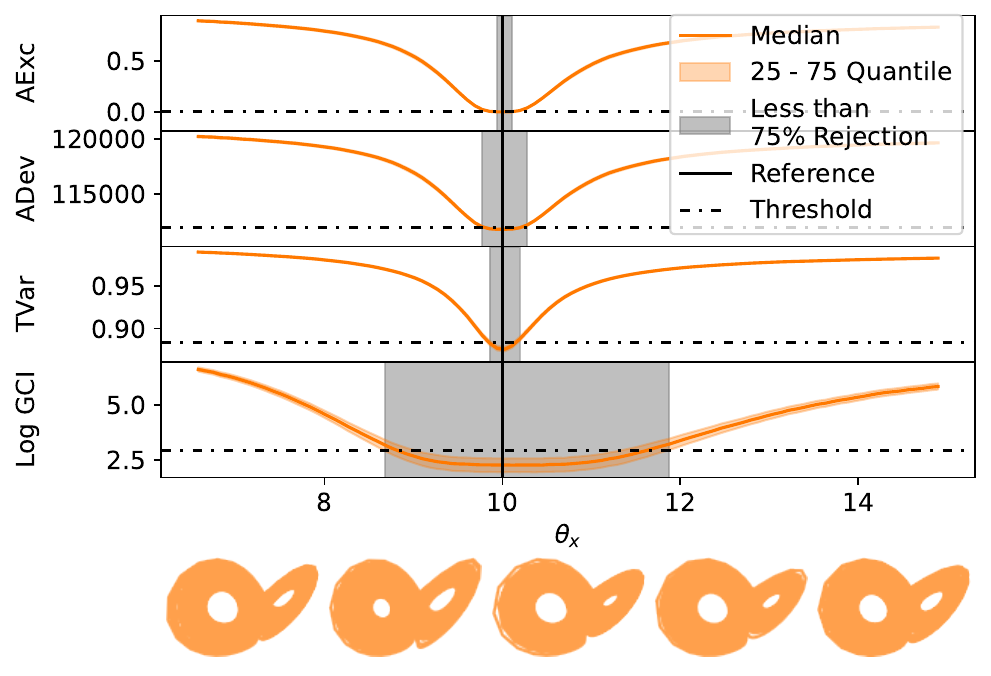}
    \caption{A one-dimensional sweep over the hyperparameter, $\theta_x$, shows that the measures capture the optimal value of $\theta_x=10$ (vertical solid black line) for Lorenz 63 with fixed $(\theta_y, \theta_z) = (28, \frac{8}{3})$. 
For each $\theta$, 1000 surrogate time series were generated with different initial conditions on the attractor to calculate the 25, 50, and 75 quantiles of the measures (orange line and area). 
The optimum is around the reference value for all measures. To capture the hyperparameter range that can be wrongly accepted by our hypothesis test, the threshold (dotted horizontal black lines) and the interval where the rejection is less than 75\% (gray area) are depicted. We used an observation interval of $\Delta t = 0.02$ and $N_O=5000$ observations for each time series. For all parameters see table \ref{tab:parameteres} upper row and for the threshold see table \ref{tab:thresholds}.}
    \label{fig:sweep_theta_02}
\end{figure}

Figure \ref{fig:sweep_theta_02} shows a one-dimensional sweep over the parameter $\theta_x$ for the Lorenz 63 system. For each value of $\theta_x$, 1000 simulations with 
$\vec \theta = \left(\theta_x, \theta_y^\text{ref}, \theta_z^\text{ref}\right)^T$
were performed to determine the test statistic values. Plotted are the 25, 50, and 75-quantiles for the GCI, TVar, ADev and AExc. The $\theta_x$ value corresponding to the reference system is shown as a red solid line. Further, the acceptance area and the threshold are given. Example trajectories of the system with different $\theta_x$ are given underneath the plots. It is hard to detect any difference between the illustrations by eye. Nevertheless, all measures show a clear, seemingly continuous, trend for their distribution with respect to $\theta_x$. All measures have a minimum near the reference value and a clear region, where at least 75\% of the systems are above the threshold. This region is significantly smaller for all measures compared to the GCI and is the smallest for the Attractor Exclusion (AExc). Based on the theoretical validation of the GCIV and the GCI, as well as the established ability of the GCIV to distinguish miss parameterized systems \cite{HAA15}, Fig.~\ref{fig:sweep_theta_02} indicates that all measures can be used to find the correct system parameter $\theta_x$ and perform at least as well as the GCI. 
\begin{figure}[h]
    \centering
    \includegraphics[width=\linewidth]{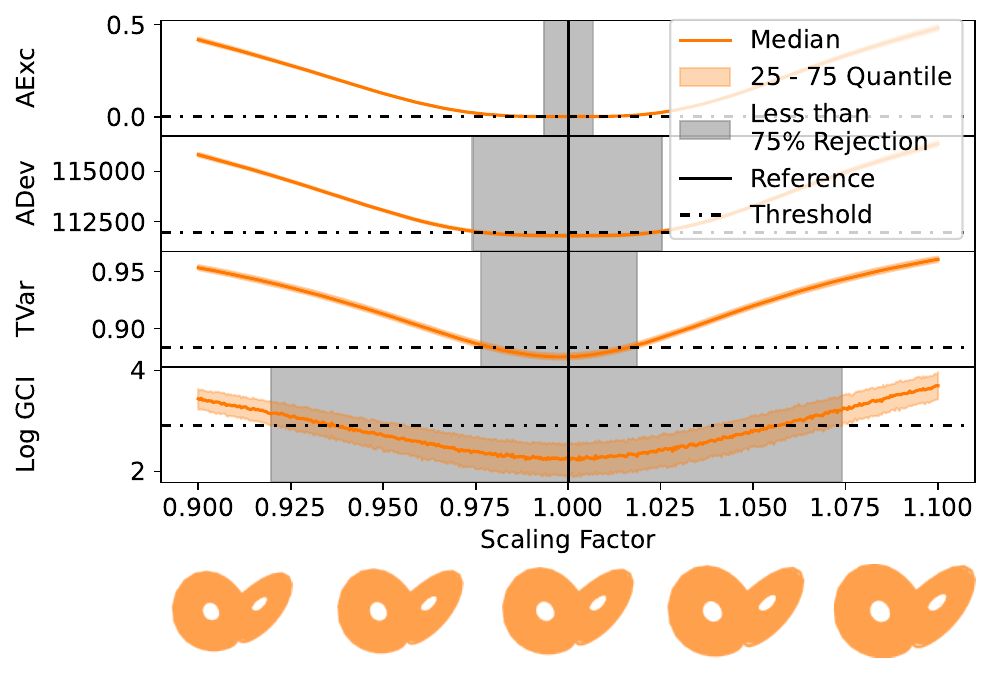}
    \caption{A one-dimensional sweep for differently scaled reference time series shows that the measures capture the optimal value of $1$ (vertical solid black line) for Lorenz 63 with fixed $(\theta_x, \theta_y, \theta_z) = (10, 28, \frac{8}{3})$.  The time series were scaled around the center of mass.
For each scaling, 1000 surrogate time series were generated from the reference system, all with different initial conditions on the attractor. From these, the 25, 50, and 75 quantiles of the measures (orange line and area) were determined. 
The optimum is around the reference value for all measures. To capture the hyperparameter range that can be wrongly accepted by our hypothesis test, the threshold (dotted horizontal black lines) and the interval where the rejection is less than 75\% (gray area) are depicted. We used an observation interval of $\Delta t = 0.02$ and $N_O=5000$ observations for each time series. For all parameters see table \ref{tab:parameteres} upper row and for the threshold see table \ref{tab:thresholds}.}
    \label{fig:sweep_shring_and_inflate_02}
\end{figure}

In Fig.~\ref{fig:sweep_shring_and_inflate_02} the test statistics are compared for test time series that have been generated by rescaling trajectories from the reference system ($\vec \theta = \vec \theta^\text{ref}$) by a small factor around its center of mass:
\begin{align}
    \vec u_t^\text{scaled} = \text{factor} \times(\vec u^\text{ref}_t - \bar {\vec u}) + \bar {\vec u}, 
\end{align}
where $\bar {\vec u}$ is the time average of the reference time series. Here, the same trends are found. The GCI appears to be the least sensitive measure for the Lorenz63, while the AExc is the most sensitive. All measures are optimal for the unaltered reference system. 



Figures \ref{fig:sweep_theta_02} and \ref{fig:sweep_shring_and_inflate_02} both  show that all measures are sensitive to small deviations from the reference systems and that the thresholds are well chosen to exclude time series that deviate from the desired dynamics. Based on these results, we can confidently use these measures to characterize the performance of surrogate models. We show examples of this in Section~\ref{sec:results} for surrogate models generated via reservoir computing. 

\subsection{Reservoir Computing}\label{sec:reservoir_computing}

Reservoir computing uses a dynamic system to transform an input time series into a higher-dimensional and dynamically-rich space. The target dynamic is reconstructed by a linear projection of the higher dimensional space into the target space. In contrast to classical machine learning, only the linear projection is trained, while the transformation of the input data is fixed and given by the dynamical system. 
Our reservoir is represented by the map 
\begin{equation}\label{eq:res_map}
\vec x_{t} \coloneq \tanh(\rho \bm W _ \text{int}\vec x _ {t-1} + k \bm W _ \text{in} \vec u ^\text{in}_{t} ),
\end{equation}
where  $\vec x_{t} \in \mathbb{R}^N$ is the response of the reservoir at time $t$, $\vec u ^\text{in}_{t} \in \mathbb{R}^N$ is the input at time $t$, $\bm W _ \text{int}=\rho \bm I$ is the coupling matrix with  spectral radius $\rho \in \mathbb{R}$ and identity matrix $\bm I \in \mathbb{R}^{N\times N}$, $ \bm W _ \text{in} \in [0, 1]^{N \times D}$ is the input matrix, $k$ is the input scaling, and $\vec x _0 = \vec 0 \in \mathbb{R}^N$ is the initial condition. Throughout this paper, the dimension of the task is $D=3$ and, unless stated otherwise, the number of nodes is $N=20$. We define different reservoirs by choosing different input matrices, where the elements are distributed uniformly between $0$ and $1$. These reservoirs correspond to the uncoupled reservoirs from \cite{JAU24a}. 

We define two phases for reservoir computing, the training phase and the autonomous phase. Each phase needs a separate time series of input samples, $\left( \vec u ^\text{train}_t \right)_{t \in 1 \dots T_\text{buf} + T_\text{train}}$ and $\left( \vec u ^\text{auto}_t \right)_{t \in 1 \dots T_\text{buf} + T_\text{auto}}$ with $T_\text{train}=10000$ and $\text{auto} = 5000$, respectively. 
We rescale each component of the time series according to
\begin{align}
    \text{normalize}(u) &= \frac{u - \min_t( u ^\text{train}_t)}{\max_t( u ^\text{train}_t) - \min_t(u ^\text{train}_t)},
\end{align}
and thus, ensuring that the training data is bounded between 0 and 1.
The time series are fed into the reservoir to get the corresponding responses of the reservoir. Furthermore, the first $T_\text{buf} = 10000$ elements of the responses are called the buffer region (washout/synchronization phase) and are discarded. This region is introduced to provide the reservoir time to forget about its past inputs and adjust to the input sequence. For the second phase, this allows us to compare the output of the reservoir's trajectory directly to the input, which is essential for calculating the valid prediction time. In the following, we consider only the data after the buffer regions. 
\begin{figure*}[t]
    \centering
    \includegraphics[width=\linewidth]{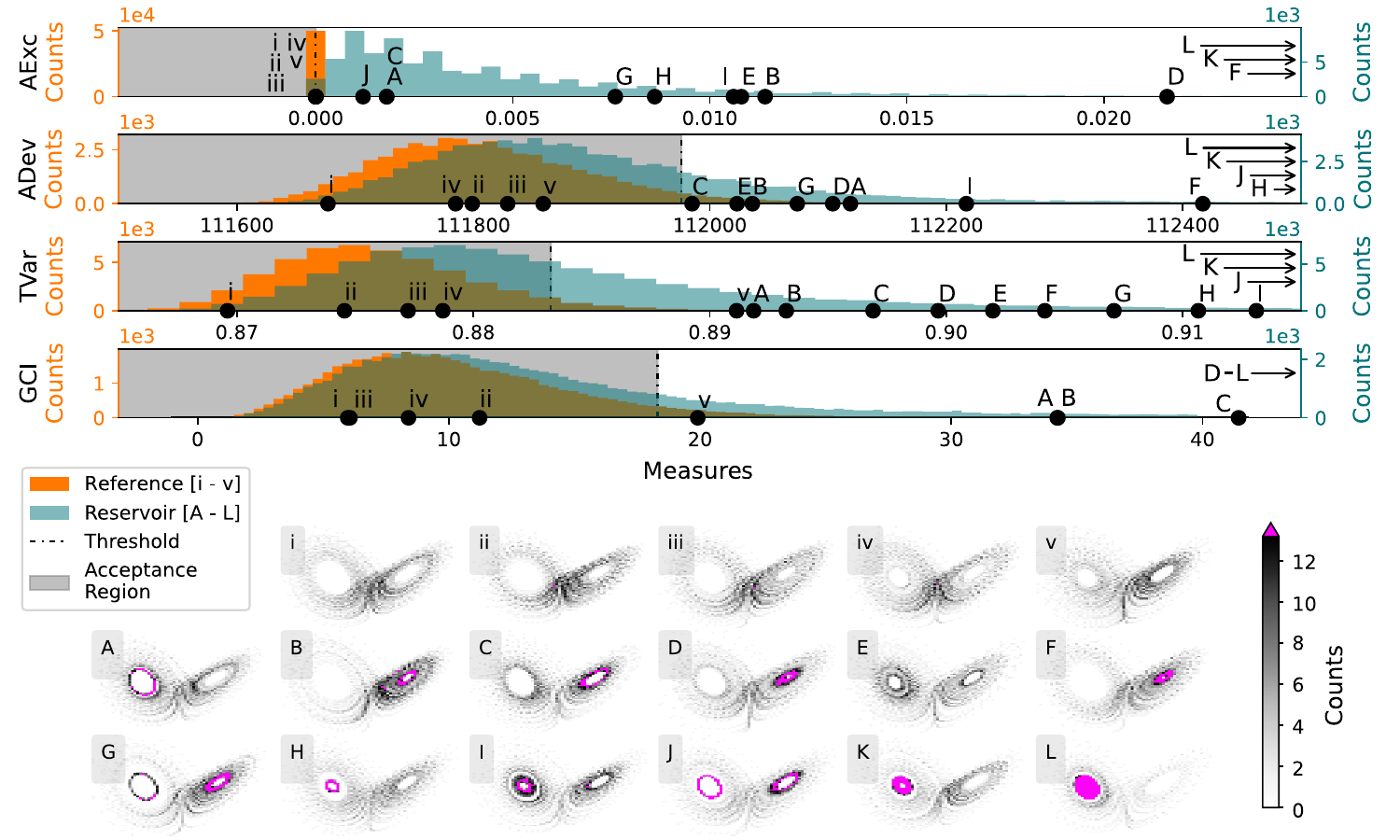}
    \caption{The turquoise (orange) histograms show the autonomous/surrogate (reference) time series evaluated for different measures. We used a fixed spectral radius of $0.4$ for the reservoir. As the reference system, we chose a Lorenz 63 system with observation time $\Delta t = 0.02$ and $N_O=5000$ observations. As an illustration of the quality of solutions, multiple trajectories are presented. 5 random trajectories were from the reference system (Roman numerals) and 12 simultaneously rejected time series from the reservoir (Latin letters). Further, the threshold (dashed vertical black line) and acceptance region (gray area) are shown to identify accepted solutions. For these trajectories, the measures are calculated and added to the histograms as labeled black dots. Arrows indicate if a value is out of range. Below the histograms, the empirical spatial probability distribution is shown for all labeled examples. For all parameters see table \ref{tab:parameteres} upper row and for the threshold see table \ref{tab:thresholds}.}
    \label{fig:application_examples}
\end{figure*}

The first phase is the training phase, where we learn the linear projection from the high-dimensional space into the target space. For the linear projection the weights, $\bm W_\text{reg}$, and the bias, $\vec b$ are obtained using ridge regression such that the target series $\vec u^{\text{trgt}}_t$ fulfills:
$\vec u^{\text{trgt}}_t \approx \bm W_\text{reg} \vec{x}_t^{train} +\bm b,\, \forall t \in K+1, K+2, \ldots$, where $\vec{x}_t^{train}$ are reservoir nodes states given by Eq.~(\ref{eq:res_map}) with input $\vec u ^\text{in}_{t}=\vec{u}^{train}_{t}$. 
In this paper, the target is the one-step-ahead prediction of the 3-dimensional input data:
\begin{align}
    \vec u^{\text{trgt}}_t \coloneq \vec{u}^{train}_{t+1}
\end{align}
The ridge regression is a linear regression with an additional L2 regularizer
\begin{align}
    \bm W_\text{reg}, \vec b &= \argmin_{ \bm W_\text{reg}, \vec b} \sum_{t=\Delta + 1}^{N} \left|\left| \vec u^{\text{trgt}}_t -  \left( \bm W_\text{reg} \vec{x}_t + \vec b \right) \right|\right|^2_2 + \nonumber \\ &+ \lambda  \left( ||\bm W_\text{reg}||^2_2 + ||\vec b||^2_2\right) ,
\end{align}
and the solution can be calculated analytically:
\begin{align}
\begin{pmatrix}
\bm W_\text{reg} \\
 \vec b^T
\end{pmatrix} &=\left( \begin{pmatrix}
\bm X^T \\
\vec{1}_N^T
\end{pmatrix}  \begin{pmatrix}
\bm X &
\vec{1}_N
\end{pmatrix} + \lambda \mathbbm{1}_{N+1}\right)^{-1} \begin{pmatrix}
\bm X^T \\
\vec{1}_N^T
\end{pmatrix} \bm U,
\end{align}
where $(\bm X)_{i, t} = (\vec x^\text{train}_t)_i$ and $(\bm U)_{i, t}= (\vec u^\text{trgt}_{t})_i \forall t \in T_\text{buf}+1, \ldots, T_\text{buf} + T_\text{train}$, and with regularizer $\lambda$.

In the final phase, the learned system is run autonomously. For this, the one-step-ahead prediction is used as the next input. This gives rise to an autonomous trained system:
\begin{equation}
    \hat{ \vec x}_{t+1} \coloneq \tanh(\rho \hat { \vec x }_ t + \bm W _ \text{in} (\bm W_\text{reg} \hat{ \vec x} _ t + \vec b) ).
\end{equation}
For the initial conditions of the autonomous system, we use the first time step after the driven (open-loop) buffer region $\hat {\vec x}_0 = \hat {\vec u}^\text{auto}_{T+1}$. 
Using the trained weights and bias to project the autonomously evolving reservoir states down to the lower dimensional space of the reference system,
\begin{equation}
    \hat{\vec u}_t = \bm W_\text{reg} \hat{\vec{x}}_t +\bm b
\end{equation}
we obtain the autonomous time series, $\left( \hat{\vec u}_{t} \right)_{t \in 1, \ldots, T_\text{auto}-1}$, on which the test statistics will be applied.
We can calculate the valid prediction time by comparing the autonomous time series to the reference time series which was used to initialize the trained reservoir, $\left( \vec u ^\text{auto}_t \right)_{t \in T_\text{buf} + 2, \ldots, T_\text{buf} + T_\text{auto}}$. The first data point at time $T_\text{buf} + 1$ is ignored, as the reservoir predicts one time step into the future. 



\section{Application}\label{sec:results}

\begin{figure*}[t]
    \centering
    \begin{subfigure}{\textwidth}
        \includegraphics[width=1.0\textwidth]{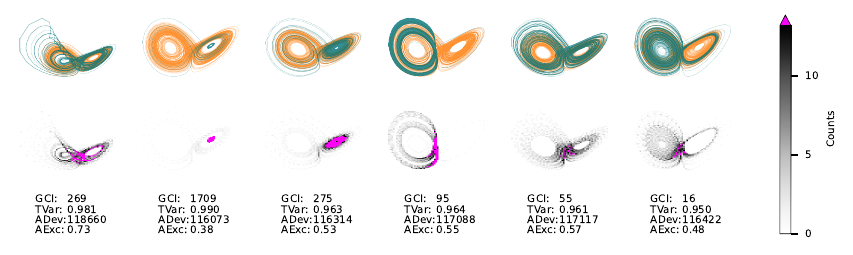}
        \caption{}
        \label{fig:problem_visualization}
    \end{subfigure}

    \begin{subfigure}{\textwidth}
        \includegraphics[width=1.0\textwidth]{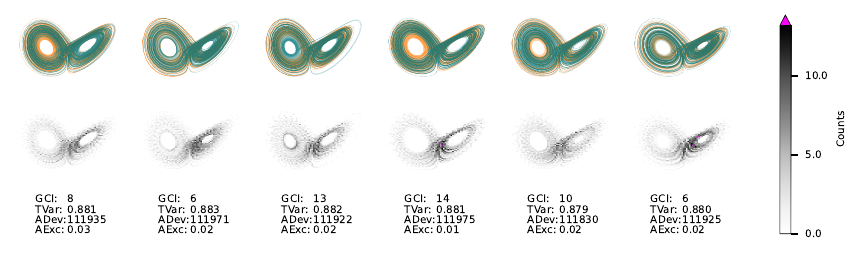}
        \caption{}
        \label{fig:good_visualization}
    \end{subfigure}

    \caption{
    In (\subref{fig:problem_visualization}) a random selection of autonomous time series from a reservoir with 10 nodes trained on a Lorenz 63. For each autonomous time series (turquoise trajectory), the respective reference time series (orange trajectory) are plotted. Below each example, the empirical spatial probability distribution is shown (gray, 2d histogram). In  (\subref{fig:good_visualization}) 6 random surrogate time series that are simultaneously accepted by the ADev, TVar, and GCI are presented. The observation time is $\Delta t = 0.02$, scaling factor $k=0.4$, spectral radius $\rho=0.4$, and $N_O=5000$ observations were generated. For this reservoir out of 50000 simulations, only 9 time series simultaneously were accepted. For all parameters see table \ref{tab:parameteres} lower row and for the threshold see table \ref{tab:thresholds}.
    }
    \label{fig:both_visualization}
    
\end{figure*}

In this section, we apply the measures introduced in Section~\ref{sec:measures} to the time series produced by the autonomous trained reservoir, $\hat{\vec u}_t$. We will show that the measures are useful tools for excluding poorly trained systems and for hyperparameter optimization.

In reservoir computing, typically, large networks of 100s to 1000s of nodes are used and the input weights, as well as the coupling weights, are chosen randomly. In our reservoir example, we have restricted the coupling to self-coupling \cite{JAU24a} and consider only N=20 nodes. Using smaller reservoirs is of interest when computational efficiency or interpretability are desired. However, smaller systems also lead to more variability in the dynamics of the final trained autonomous surrogate systems. As described in \cite{JAU24a} the dynamics can be roughly sorted into three categories according to whether they remain oscillatory throughout the entire autonomous run, collapse to a fixed point, or diverge past a certain threshold. However, within the category of those systems that remain bounded and oscillatory, there is still a need to distinguish between those which behave like the reference system and those which do not. Therefore, in the following we will concentrate on applying the measures to those autonomous test time series for which the absolute value of each coordinate is bounded by 2, i.e. $\max_{t\in 1, \ldots, T}(\|\vec u^\text{auto}_t\|_\text{max}) < 2$, where $\|\vec u\|_\text{max} = \max_i{|u_i|}$, and the three last values of a time series fluctuate more than $5\times 10^{-3}$
, i.e. $\min_{t \in \{2, 1\}}(\|\vec u^\text{auto}_{T_\text{auto}-t} - \vec u^\text{auto}_{T_\text{auto}} \|_\text{min}) > 5 \times 10^{-3}$, where $\|\vec u\|_\text{min} = \min_i{|u_i|}$. Both definitions are equivalent to those used in \cite{JAU24a}. This category of surrogate systems also encapsulates those for which the largest conditional Lyapunov is significantly smaller than the Lyapunov exponents of the reference system, a condition that was related to successful attractor reconstruction in \cite{HAR24}.

Figure \ref{fig:application_examples} shows histograms of the values of the four measures for different realizations of the input time series and input weights for a fixed spectral radius of $\rho = 0.4$, with the histogram for the reference shown in orange. Each of the measures is designed such that a lower value is better, therefore all values below the threshold are accepted (Acceptance Region in grey). If a sample is larger than the threshold, it is considered rejected. For the AExc almost all samples are rejected, indicating this measure is too exclusive to be useful, while for the other measures at least 26\% are rejected for this choice of reservoir. 
Plotted below the histograms are projections of the empirical spatial probability density function of reference and test time series. We included 5 random samples from the reference system (Roman numerals) and 12 samples that were rejected by all measures simultaneously (Latin letters). The measures capture different properties of the reference system, therefore disagreement between the measures is possible. In this case, only 15\% of the surrogate systems were simultaneously rejected by all test statistics.
We see a clear visual distinction between the rejected and reference time series, which gives us confidence that the measures correctly sort out bad reconstructions. 
%
%
%

When the size of the reservoir network is further reduced, the dynamics of the trained autonomous systems can vary sufficiently from the reference system such that they can be excluded by visual inspection. Ideally, the measures would exclude all such cases. In Fig.~\ref{fig:problem_visualization} six examples of such systems are shown for reservoirs of size $N=10$. Shown are projections of the trajectories and corresponding empirical spatial probability densities, along with the calculated values of the four measures. For all except the rightmost example, all measures are above the respective thresholds. However, according to the GCI value the rightmost one would not be excluded. That such cases can occur is not unexpected, since the thresholds allow statements to be made about the likelihood of the systems not behaving like the reference system, but not about the likelihood of the systems behaving like the reference system. 

To mitigate the false acceptance of surrogate systems one can use a combination of these measures. Ideally, if we set an overall acceptance criteria that a surrogate system must be accepted simultaneously by the GCI, TVar, and ADev, then all accepted systems should also pass a visual inspection. Here we do not use the AExc because the previous results have shown that it is too exclusive to be useful in this context. Using the $N=10$ reservoir that was also used for Fig.~\ref{fig:problem_visualization}, we find that surrogate systems that are simultaneously in acceptance regions of the GCI, TVar, and ADev do in general lead to trajectories that cannot be discarded via visual inspection. In Fig. \ref{fig:good_visualization} six examples of randomly chosen autonomous surrogate systems that pass the overall acceptance criteria are depicted. For this demonstration a poorly performing reservoir setup was chosen, hence, only 9 of 50000 reservoirs fulfill our criteria. These results indicate that combining the measures captures enough properties to mostly include only visually accurate solutions.

We used hypothesis testing to reduce our set of trained surrogate systems to those that are likely to exhibit the dynamics of the reference system. Although it would appear that the measures cannot be used to rank surrogate time series due to the inherent distribution of test statistic, we showed in section \ref{sec:validation} that the variation of different realizations is less than the overall trend. Thus, it appears possible to compare samples, if they are far enough apart, with high confidence. This leads to the idea to use the measure as an optimization tool.



\begin{figure}[t!]
    \centering
    \includegraphics[width=\linewidth]{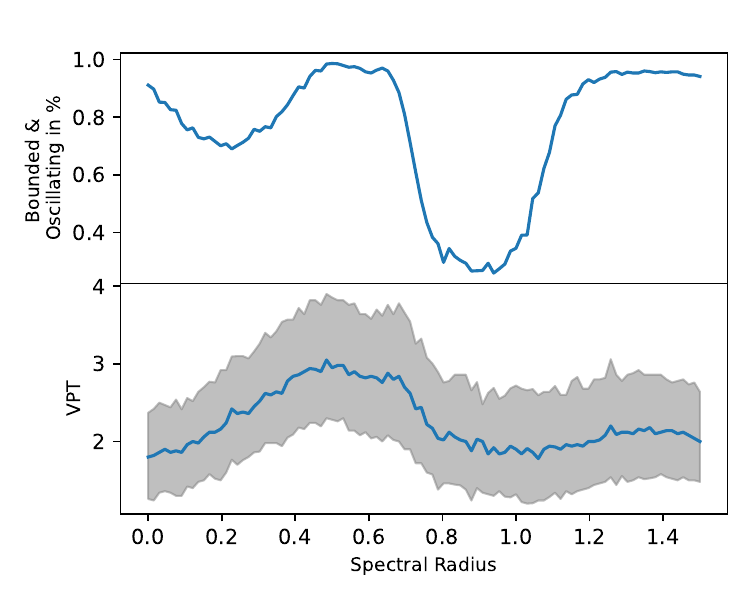}
    \caption{Valid Prediction Time (VPT) and the percentage of bounded and oscillating solutions for autonomous trajectories from a 20-node Reservoir. For each spectral radius, we used 1000 reservoirs with different input weights and training data. For the VPT the 25, 50, and 75 quantiles are plotted (blue line and gray area). As higher values are better, we find an optimum around 0.5 for both. The data was generated using a Lorenz 63 with an observation time of $\Delta t =0.02$. For all parameters see table \ref{tab:parameteres} upper row.}
    \label{fig:application_vpt_vs_spectral_radius}
\end{figure}

\begin{figure*}[t!]
    \centering
    \includegraphics[width=\textwidth]{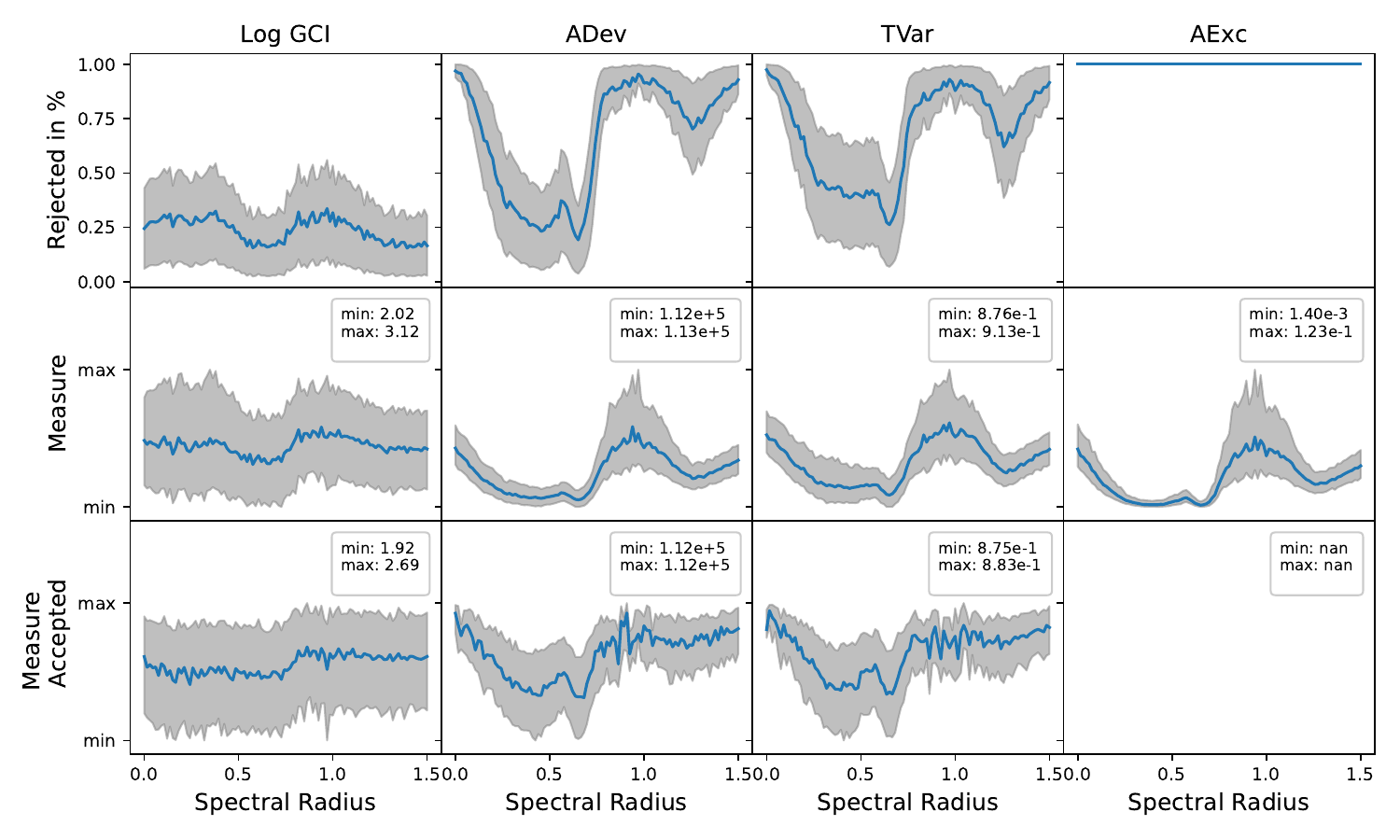}
    
    \caption{Impact of spectral radius for reservoirs with 20 nodes trained on the Lorenz 63. For each spectral radius, we simulated 1000 reservoirs with different input weights and input data to generate the autonomous trajectories. For all bounded and oscillating trajectories, we summarize the measures 25, 50, and 75 quantiles (middle row; blue line and gray area). To see how many samples are significantly different from the reference system, the top row of the figure shows the percentage of rejected trajectories using the hypothesis test. Further to evaluate the quality of the accepted solutions, the 25, 50, and 75 quantiles of only the accepted trajectories are shown in the bottom row. As lower values are better, the optimal value agrees across the measures to be around a spectral value of $0.6$. In contrast to the other measures, AExc accepts no sample. We used an observation time of $\Delta t =0.02$. For all parameters see table \ref{tab:parameteres} upper row and for the threshold see table \ref{tab:thresholds}.}
    \label{fig:application_measures_vs_spectral_radius}
\end{figure*}

We want to investigate the usefulness of the measures for hyperparameter tuning. For this, we consider the impact of the spectral radius $\rho$, an important hyperparameter in reservoir computing \cite{VER07,HAL19,VIE23,JAU24a}. In Figure \ref{fig:application_vpt_vs_spectral_radius} and \ref{fig:application_measures_vs_spectral_radius}, we look at different spectral radii between $0$ and $1.5$. Analogous to the analysis in \cite{JAU24a}, in Fig.~\ref{fig:application_vpt_vs_spectral_radius} the valid prediction time and the percentage of bounded and oscillating solutions are calculated for all realizations of the reservoir. The best performance in terms of the VPT occurs near a spectral radius of $\rho=0.5$ and coincides with a maximum in the percentage of realization which are bounded and oscillating over the entire autonomous prediction run. Complementary to these results, results relating to the four measures are shown in Fig.~\ref{fig:application_measures_vs_spectral_radius}. We remind the reader that the measures are only evaluated for the reservoir realizations which result in bounded and oscillating time series. The percentage of bounded and oscillating solutions that are rejected by the measures are shown in the top row of \ref{fig:application_measures_vs_spectral_radius}. Of these bounded and oscillating time series, the middle row shows the values of the measures and the bottom row shows the values for the subset of test trajectories that were accepted by the respective thresholds. In each case, the blue curve indicates the 50$^{th}$ quantile of the distribution of values, with the ranges to the 25$^{th}$ and 75$^{th}$ quantiles indicated in gray.

We find that the rejection rate of all measures captures an optimum around the spectral radius of $0.6$. Although the VPT measures a local property, rather than our measures that target a global property of the attractor, the optima are close together.
The GCI has a less pronounced trend and rejects less than the other measures. 
The AExc rejects almost all samples regardless of the spectral radius. 

As discussed in the validation section, we showed that the measures capture the optimal value and that smaller values are an indication of the proximity to the optimal value, even though this is not theoretically justified. This leads to the idea of using the measures as a distance measure directly. We find that the distribution of the AExc values shows similar trends as the ADev and TVar, as seen in Fig.~\ref{fig:application_measures_vs_spectral_radius}(middle row), indicating some comparative information can be extracted from these measures. Again, the GCI has a less pronounced trend, however, it captures an optimum for a similar spectral radius.

If we remove all the rejected time series and analyze the remaining ones, we discover the same trend as before (see bottom row of Figure \ref{fig:application_measures_vs_spectral_radius}). Indicating that for the optimal value not only fewer time series are rejected, but also that the overall quality of the reconstruction is improving. Similar results are shown for other chaotic reference systems in the supplemental material.

\section{Discussion/Conclusion}


%

In this paper, we have introduced four measures that use global properties of the attractor for characterizing chaotic trajectories. In contrast to previous methods, the evaluation of the measures does not require any fitting processes, and is thus straightforward, while still being closely related to global properties of the attractor. The Generalized Correlation Integral (GCI) and the other measures are based on the correlation integral and the spatial probability density function, respectively, and thus, capture different properties of the attractor. The GCI is based on the already established first version of the Generalized Correlation Integral Vector\cite{HAA15}, and thus, was used as a benchmark. We applied the hypothesis testing framework to categorize samples if they are significantly different from the true system using a simple rejection criterion. In the validation section, we showed empirically that all of the measures (GCI, TVar, ADev, and AExc) can be used for system parameter identification.

We applied the measures to reconstructed time series from a reservoir, i.e. the autonomous time series. With 20 nodes, our reservoir has a small number of nodes compared to common reservoir implementations \cite{HAR24,PAT18a,RAT25,VIE23,FLY23}. For such a small number of nodes, the dynamics of the autonomous time series can vary significantly\cite{JAU24a}. In \cite{JAU24a}, the author searched for the spectral radius, $\rho$, to find the optimal reservoir using the ADev. We showed that the GCI, AExc, and TVar can be used alongside the ADev for hyperparameter optimization of the spectral radius. Using the hypothesis testing scheme, we were able to exclude bad reconstruction systematically. When using the measures as rank statistics, for the optimal spectral radius, the reservoir is less likely to be rejected and the quality of the overall solutions improves.
Further, we showed for an uncoupled reservoir with 10 nodes if we look simultaneously at accepted values under the measures GCI, TVar, and ADev they appear to be visually good examples. Although the hypothesis testing approach gives no theoretical guarantee that accepted samples are from the true system, our findings let us believe that the rate of falsely accepted reservoirs is low. Our results demonstrate the advantage of including the statistical nature of measures applied to chaotic systems in a meaningful manner. Using our approach new insights can be gained into methods of generating accurate and interpretable surrogate models.

Although our measures capture a global quantity of the attractor, we disregard the temporal information of the model. Further investigations have to be done into methods that capture the temporal dynamic as well as how they can be combined with our methods. A promising direction would be to develop test statistics based on the probability distribution of the spectra of the reference system. Such an approach could be robust against sparse temporal sampling, given a long enough test time series.


    


\FloatBarrier

\section{Supplementary Material}

 In the supplemental material, we show results for four additional low-dimensional chaotic systems, namely Sprott B, Sprott C, Sprott D\cite{SPR94}, and Roessler76\cite{ROE76}. In addition, for each system (including the Lorenz63), we provide the results for a short observation time, $\Delta t$, and long observation time, $\Delta t_\text{Sparse} = 50 \times \Delta t$. We will refer to systems with long observation times as the sparse (observation) systems. Further, we include the results for the PMCMC sampler (analog to figure \ref{fig:heikki_theta_given_y}) using the original algorithm\cite{HAA15}.

\section{Acknowledgments}

This work was 
made possible by funding from the Carl-Zeiss-Stiftung.

The research of Jana de Wiljes has been partially funded by the Deutsche Forschungsgemeinschaft (DFG)- Project-ID 318763901 - SFB1294.
Furthermore this project has received funding from the European Union under the Horizon Europe Research \& Innovation Programme (Grant Agreement no. No 101188131 UrbanAIR). Views and opinions expressed are however those of author(s) only and do not necessarily reflect those of the European Union. Neither the European Union nor the granting authority can be held responsible for them.

\section{Data Availability Statement}

The data that support the findings of this study are openly available in Data-Driven Performance Measures using Global Properties of Attractors for Black-Box Surrogate Models of Chaotic Systems Repository at \url{http://doi.org/10.5281/zenodo.15564346}. 

\appendix

\section{Thresholds}

\begin{table}[H]
    \centering
    \begin{tabular}{|l|r|r|r|}
\hline
            AExc &   ADev &     TVar &     GCI \\
\hline
 0 &  86465 & 0.848 & 18.67 \\\hline
\end{tabular}
\caption{Empirically determined thresholds for different Measures for the Lorenz63 systems by repeatably measuring $50000$ reference time series with an observation time of $\Delta t = 0.02$ and $5000$ observations per time series. The threshold is determined as the 0.95-quantile of the resulting distribution. In the paper, all measures except the GCI use this empirical threshold for the hypothesis testing. A theoretical value of $18.307$ is used for the GCI since the distribution is known. By definition, all values of the AExc are 0, and thus, the threshold as well.}
\label{tab:thresholds}
\end{table}

\section{GCI - Miss Parametrization}\label{App:GCI}

The General Correlation Integral (GCI) measure is able to differentiate miss parameterized systems from the reference system for a wide variety of systems \cite{SPR19, HAA15}. In the following, we will restrict ourselves to the General Correlation Integral Vector (GCIV) as it is directly linked to the GCI. In this section, we will introduce a novel interpretation of this previous method using similarity measures for probability density functions. 

First, we will switch to a probabilistic framework, where the GCIV, $\vec y$, will be seen as a Random variable, $Z$, depending on the random variables of the initial position of the surrogate system $X$ and the unknown (and thus modeled as random) parameter of the surrogate system, $\Theta$,

\begin{align}
    Z = \vec y(X, \Theta),
\end{align}

i.e. we have the probability density ${p}_{Z| X, \Theta}(\vec z|\vec x, \vec \theta) = \delta(\vec z - \vec y(\vec x, \vec \theta)) $ given, where $\delta$ is the Dirac delta distribution. W.l.o.g. the initial position of the reference system $X^\text{ref}$ is ignored as it is treated the same as $X$. We assume that ${p}_ {X|  \Theta}$ is known and sampleable, in our case it is chosen to be a point on the attractor. Thus, we can express

\begin{align}
    {p}_{Z | \Theta}(\vec z | \vec \theta) &= \int {p}_{Z, X | \Theta}(\vec z, \vec x | \vec \theta) d \vec x \nonumber\\
    &= \int {p}_{Z | X, \Theta}(\vec z| \vec x, \vec \theta) \times  {p}_{X | \Theta}(\vec x | \vec \theta) d \vec x\nonumber \\
    &= \int \delta(\vec z - \vec y(\vec x, \vec \theta)) \times  {p}_{X | \Theta}(\vec x | \vec \theta) d \vec x.
    \label{eq:apdx:heikki:p_z_given_theta}
\end{align}

When generating samples from a marginal distribution, it is sufficient to sample from the corresponding joint distribution and apply the appropriate transformation. Thus, to sample from ${p}_{Z | \Theta}(\vec z | \vec \theta)$ it is sufficient to sample from the integrant, $\delta(\vec z - \vec y(\vec x, \vec \theta)) \times  {p}_{X | \Theta}(\vec x | \vec \theta)$, which is possible as the expression for $\vec y$ is known and ${p}_ {X|  \Theta}$ is assumed to be samplable.

Second, we express the previous paragraphs in our probabilistic framework, where we calculated $\bm \Sigma$ and $\vec{\mu}$. They can be seen as approximating the marginal distribution of $Z$ given $\vec \theta^\text{ref}$ as a normal distribution (see algorithm \ref{alg:heikki_ref})

\begin{align}
    p_{Z | \Theta}(\vec z | \vec \theta^\text{ref} ) &\stackrel{\ref{eq:apdx:heikki:p_z_given_theta}}{=} \int \delta(\vec z - \vec y(\vec x, \vec \theta_0)) \times  {p}_{X | \Theta}(\vec x | \vec \theta_0) d \vec x
    \nonumber\\&\approx \mathcal{N}(\vec z; \vec \mu, \bm \Sigma),
\end{align}

where the influence of the surrogate initial conditions was marginalized. This marginal distribution is our likelihood function for the reference system, as it represents the likelihood for $\vec y$ if the surrogate time series is a realization from the reference system.

The previous methods used algorithm \ref{alg:heikki_mcmc} to sample from the unnormalized posterior distribution, 
 \begin{align}
     \pi(\vec \theta) &= \int  p_{Z | \Theta}(\vec y(\vec x, \vec\theta) | \vec \theta^\text{ref} ) \times p_{X | \Theta}(\vec x | \vec \theta) d \vec x.
     \label{eq:apdx:heikki:target}
 \end{align} 
 It was shown that algorithm \ref{alg:heikki_mcmc} is a Pseudo-Marginal Markov Chain Monte Carlo (PMCMC) algorithm sampling from $\pi(\vec\theta)$\cite{HAA15}. 



\begin{figure*}[th!]
    \centering
    \begin{minipage}[t]{0.95\textwidth} 
    \includegraphics[width=\linewidth]{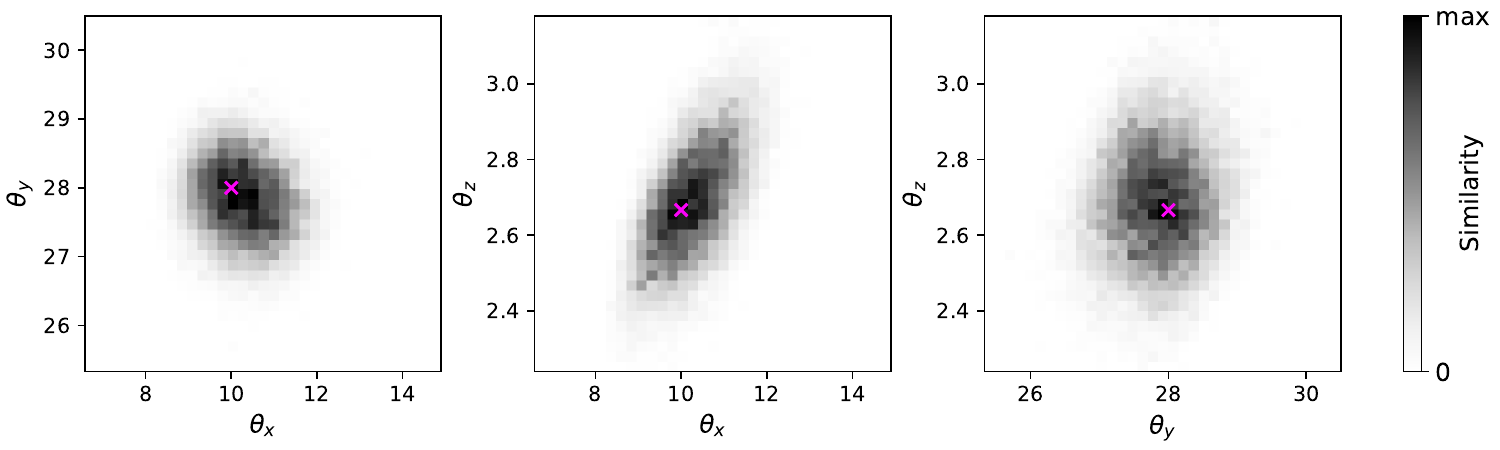}
    \end{minipage}
    \caption{Shows marginal for the similarity measure. The similarity is measured using the L2 inner product of the correlation integral vector distribution for the reference system and a system that is parameterized with $(\theta_x, \theta_y, \theta_z)$. The pink cross indicates the parameters of the reference system. The maximum value from left to right is 7891, 7107, and 5354. We use different-sized reconstruction and reference time series in contrast to \cite{HAA15} to make the posterior narrower. Changes in the transition kernel variance $\Sigma_F$ had no major influence. We used the Lorenz63 with an observation time $\Delta t = 0.02$. For all parameters see table \ref{tab:parameteres} upper row and table \ref{tab:mcmc}}
    \label{fig:heikki_theta_given_y}
\end{figure*}

Finally, we show that the probability $\pi(\vec\theta)$ can be seen as measuring similarity. For that, we introduce a similarity measure. If we assume that the random variables do have a probability density function and that they are in L2, which is justified for the reference system as $p_{Z|\Theta}$ is Gaussian, we can use the scalar product of L2 as a similarity measure

\begin{align}
    \left< q(\vec z), p(\vec z) \right>_{\vec z } = \int_{\mathbb{R}^R} q(\vec z) p(\vec z) d \vec z.
    \label{eq:apdx:heikki:similarity}
\end{align}

The unnormalized target distribution can be rewritten as
 \begin{align}
     \pi(\vec \theta) &\stackrel{\ref{eq:apdx:heikki:target}}{=}  \int  p_{Z | \Theta}(\vec y(\vec x, \vec\theta) | \vec \theta^\text{ref} ) \times p_{X | \Theta}(\vec x | \vec \theta) d \vec x \nonumber\\
     &= \iint  p_{Z | \Theta}(\vec z | \vec \theta^\text{ref} ) \times \delta(\vec z - \vec y(\vec x, \vec \theta)) d \vec z \times p_{X | \Theta}(\vec x | \vec \theta) d \vec x \nonumber\\
     &=  \int  p_{Z | \Theta}(\vec z | \vec \theta^\text{ref} ) \int \delta(\vec z - \vec y(\vec x, \vec \theta)) \times p_{X | \Theta}(\vec x | \vec \theta)  d \vec x d \vec z \nonumber\\
     &\stackrel{\ref{eq:apdx:heikki:p_z_given_theta}}{=}  \int  p_{Z | \Theta}(\vec z | \vec \theta^\text{ref} ) \times p_{Z | \Theta}(\vec z | \vec \theta )  d \vec z \nonumber\\
     &\stackrel{\ref{eq:apdx:heikki:similarity}}{=} \left< p_{Z | \Theta}(\vec z | \vec \theta^\text{ref} ), p_{Z | \Theta}(\vec z | \vec \theta )  \right>_{\vec z}.
 \end{align} 
 Thus, by sampling $\pi(\vec \theta)$ we calculate the similarity between the likelihood of the GCIV for the reference and for the miss-parameterized system using $\vec \theta$. 

 In our final results, we use the adaptive metropolis algorithm \cite{HAA01}. We choose $\Sigma_F = 0.01 \times \mathbbm{1}$, $s_d = 2.4^2 / 3$, $t_0 = 50$, and $\epsilon = 0$. We simulated additional smaller and larger  $\Sigma_F$ without any major changes.
$p_{X|\Theta}$ is defined as the probability for a random point on the attractor of the miss parameterized system. For that, we start the simulation close to the attractor and let it run for 10000 time steps. The resulting point is taken as the initial condition for the MCMC algorithm, which assumes ergodicity and an attractor. 

Figure \ref{fig:heikki_theta_given_y} shows the similarity distribution, $\pi(\vec\theta)$ for the different parameters of the Lorenz63 system, as well as a realization of the system underneath. Any parameter in the blue area is significantly different from the reference system. Thus, the GCI test statistic is able to sharply distinguish different miss parametrized models from the reference model. In contrast to \cite{HAA15}, where the algorithm uses surrogate and reference time series of the same length, in our method the similarity spread is reduced (see supplementary).





\section{Parameters}

Simulation and model  parameters are given in Table II and Table III. 

\begin{table*}[htb!]
\begin{tabular}{|l|l|l|l|l|l|l|l|l|l|l|l|}
\cline{1-10}
Figure                                                                                                                                                                                                                                                                                                                                                                                                                                                                                                                                                        & Model                                                                                                    & \begin{tabular}[c]{@{}l@{}}Number of\\ Observations, \\ $N_O$\end{tabular} & \begin{tabular}[c]{@{}l@{}}Integration \\ Method\end{tabular}   & \begin{tabular}[c]{@{}l@{}}Observation \\ Interval, $\Delta t$\end{tabular} & \begin{tabular}[c]{@{}l@{}}Integration \\ invernal, $h$\end{tabular} & \begin{tabular}[c]{@{}l@{}}Nodes, \\ $N$\end{tabular} & \begin{tabular}[c]{@{}l@{}}Input \\ Scaling, $k$\end{tabular} & Regulizer          & \begin{tabular}[c]{@{}l@{}}GCI \\ Subsample- \\rate, $c_\text{sub}$\end{tabular}  \\ \cline{1-10}
 \begin{tabular}[c]{@{}l@{}}\ref{fig:heikki_chi2} - \ref{fig:application_examples}, \\ \ref{fig:application_vpt_vs_spectral_radius} - \ref{fig:heikki_theta_given_y} \end{tabular} & Lorenz63                                                                                                 & 5000                                                                    & \begin{tabular}[c]{@{}l@{}}scipy.integrate\\.RK45\cite{scipy}\end{tabular} & 0.02                                                                        & 0.001                                                                & 20                                                    & 0.4                                                           & $1 \times 10^{-7}$ & 1/15 \\ \cline{1-10}
 \ref{fig:both_visualization}, \ref{fig:examples_lorenz}, \ref{fig:examples_reservoir}                                                                                                                                                                                                                                                                                                                                                                                                                                                                                                       & Lorenz63                                                                                                 & 5000                                                                    & \begin{tabular}[c]{@{}l@{}}scipy.integrate\\.RK45\cite{scipy}\end{tabular} & 0.02                                                                        & 0.001                                                                & 10                                                    & 0.4                                                           & 0                  & not used \\ \cline{1-10}
\end{tabular}
\caption{Hyperparameters of the Solver, Reservoir, and GCI Measures.}
\label{tab:parameteres}
\end{table*}

\begin{table*}[htp!]
\begin{tabular}{|l|l|l|l|l|l|}
\hline
Figure                                                                      & Model                                                                                                                                     & $\Sigma_F$ & $s_d$       & $t_0$ & $\epsilon$ \\ \hline
\begin{tabular}[c]{@{}l@{}}\ref{fig:heikki_theta_given_y}\end{tabular} & \begin{tabular}[c]{@{}l@{}}Lorenz63\end{tabular} & $ 0.01 \times\mathbbm{1}$     & $2.4^2 / 3$ & $50$  & $0$        \\ \hline
\end{tabular}
\caption{Hyperparameters of the Adaptive Metropolis Hastings sampler.}
\label{tab:mcmc}
\end{table*}

\FloatBarrier

\section{Algorithms}

\begin{algorithm}[H]
\caption{Create reference dataset for GCI. 
}\label{alg:heikki_ref}
\begin{algorithmic}[1]
\For{$i \in 1, \ldots, 300$} 
\State sample initial condition $\vec u_0 \sim \mathbb{P}(X|\Theta=\theta_0)$
\State simulate trajectory $U = (\vec u_t)_{t \in 1, \ldots, N}$
\State $A_i\gets U$
\State
\State sample initial condition $\vec u_0 \sim \mathbb{P}(X|\Theta=\theta_0)$
\State simulate trajectory $U^\text{ref} = (\vec u_t)_{t \in 1, \ldots, N^\text{ref}}$
\State $B_i\gets U^\text{ref}$
\EndFor
\For{$i \neq j; i, j \in 1, 2, \ldots, 300$}
    \State calculate $\vec y_{i,j} \gets \vec y(A_i, B_j)$ 
\EndFor
\State $\vec{\mu} \gets \frac{1}{300^2} \sum_{i,j=1}^{300} \vec y_{i,j}$
\State $\bm \Sigma \gets \frac{1}{300^2} \sum_{i, j=1}^{300} ( \vec y_{i,j} - \vec{\mu})^2$
\end{algorithmic}
\end{algorithm}

\begin{algorithm}[H]
\caption{PMCMC Sampler Pseudo Code}\label{alg:heikki_mcmc}
\begin{algorithmic}[1]
\For{$i \in 1, 2, \ldots, M$}
\State draw proposal $\vec \theta^p \sim \mathcal{N}(\cdot; \vec \theta, \bm \Sigma)$
\State draw proposal $\vec u_0^\text{ref} \sim \mathbb{P}(X|\vec \theta)$ 
\State draw proposal ${ \vec u}_0^p \sim \mathbb{P}(X| \hat { \vec \theta})$ 
\State Let simulation evolve that the initial condition is on the attractor  $\vec u^\text{ref}_0, {\vec u}^p_0$
\State Simulate surrogate trajectory $U^p = (\vec u_t^p)_{t \in 1, \ldots, N}$ with $\vec \theta^p$
\State Simulate reference trajectory $U^\text{ref} = (\vec u_t^\text{ref})_{t \in 1, \ldots, N}$ with $\vec \theta^\text{ref}$
\State calculate proposal $\vec y^p \gets \vec y(U^p, U^\text{ref})$
\State $\alpha \gets \min\left(1,  {\mathcal{N}(\vec y^p; \vec{\mu}, \bm \Sigma)} / {\mathcal{N}(\vec y_{i-1}; \vec{\mu}, \bm \Sigma)} \right) $
\State {\textbf{with} probability $\alpha$} \textbf{do}
    $(\vec y_i, \vec \theta_i) \gets (\vec y^p, \vec \theta^p)$
\State {\textbf{else with} probability $1-\alpha$} \textbf{do}
    $(\vec y_i, \vec \theta_i) \gets (\vec y_{i-1},\vec \theta_{i-1})$
\EndFor
\State \textbf{return} $\left(\vec \theta_i\right)_{i\in1, \ldots, M} $
\end{algorithmic}
\end{algorithm}

\section{Spatial probability distributions}

\begin{figure*}[hbt!]
    \centering
    \includegraphics[width=\linewidth]{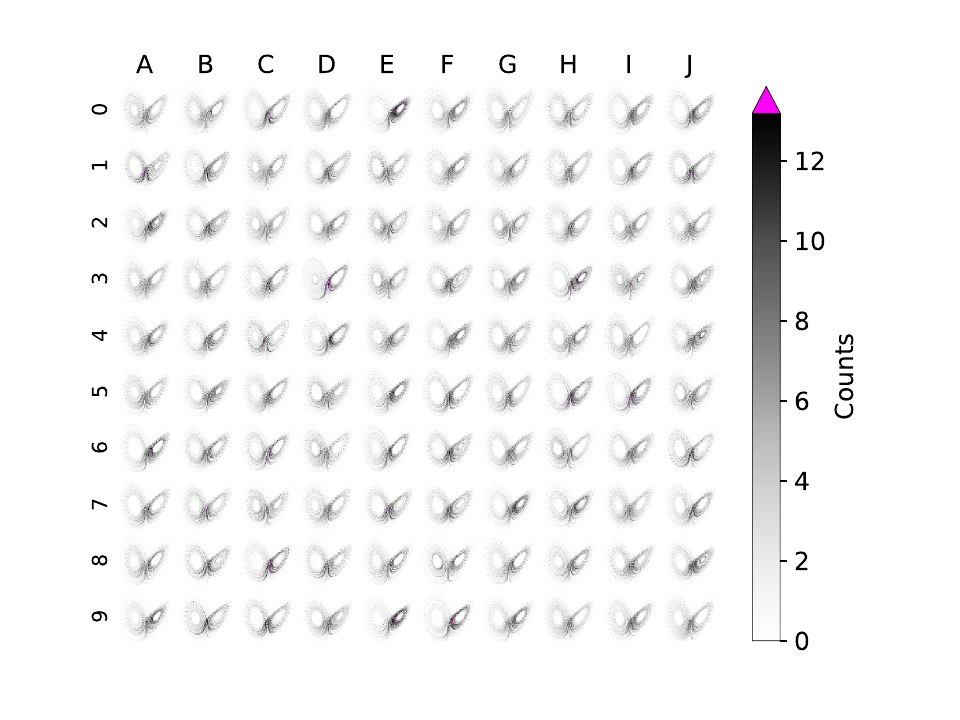}
    \caption{
    A random 10 by 10 selection of reference time series from a Lorenz63 system. For each time series the empirical spatial probability distribution is shown to build an intuition for the reference system for figure \ref{fig:both_visualization}.  For all parameters see table \ref{tab:parameteres} lower row.}
    \label{fig:examples_lorenz}
\end{figure*}

Figure \ref{fig:examples_lorenz} shows the spatial probability distribution of trajectories generated by the Lorenz63 system. This figure is a visualization of the variation that can rise in the reference system.

\begin{figure*}[hbt!]
    \centering
    \includegraphics[width=\linewidth]{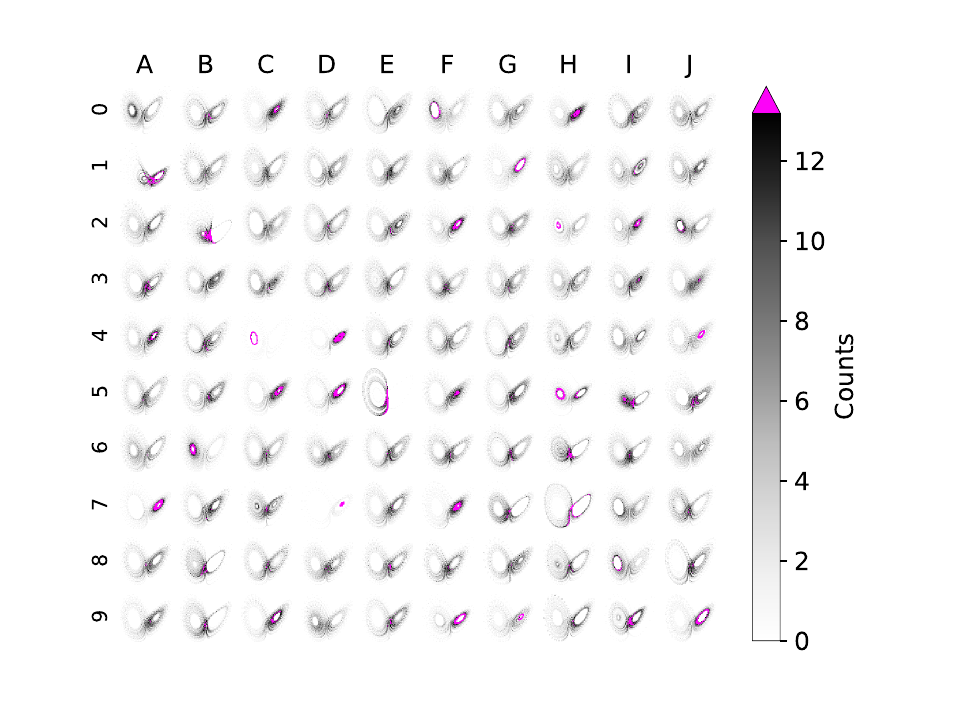}
    \caption{A random 10 by 10 selection of bounded and oscillating autonomous time series using a reservoir with 10 nodes trained on a Lorenz63. For each autonomous time series the empirical spatial probability distribution is shown, to give further examples to figure  \ref{fig:problem_visualization}. The observation time is $\Delta t = 0.02$, with input scaling $k=0.4$, and spectral radius $\rho = 0.4$. For all parameters see table \ref{tab:parameteres} lower row.}
    \label{fig:examples_reservoir}
\end{figure*}

Figure \ref{fig:examples_reservoir} shows spatial probability distributions for 100 randomly chosen surrogate models generated from the 10 node reservoir.  

\FloatBarrier

\section*{References}
\bibliography{agluedgebib_export_2025_05_30, references}

\end{document}



\title{Supplemental material: Data-Driven Performance Measures using Global Properties of Attractors for Black-Box Surrogate Models of Chaotic Systems} 



\author{L. Fumagalli}
\email[]{luci.fumagalli@tu-ilmenau.de}
\affiliation{Technische Universität Ilmenau, Institute of Physics}

\author{K. Lüdge}
\affiliation{Technische Universität Ilmenau, Institute of Physics}

\author{J. de Wiljes}
\affiliation{Technische Universität Ilmenau, Institute of Mathematics}
\affiliation{Lappeenranta University of Technology, Institude of Mathematics}
\author{H. Haario}
\affiliation{Lappeenranta University of Technology, Institude of Mathematics}

\author{L. Jaurigue}
\affiliation{Technische Universität Ilmenau, Institute of Physics}


\date{\today}

\pacs{}

\maketitle 

\tableofcontents

\newpage

\section{GCI - Miss Parameterization}\label{App:GCI}

\begin{figure}[H]
    \centering
    \includegraphics[width=\linewidth]{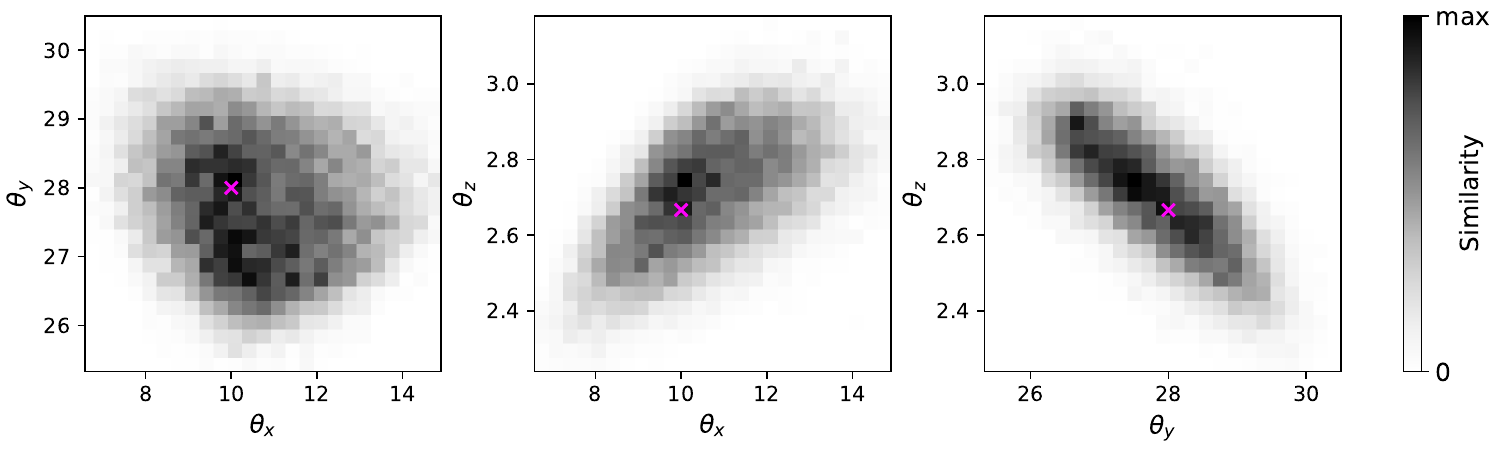}
    \hfill
    \caption{Shows marginal for the similarity measure. The similarity is measured using the L2 inner product of the correlation integral vector distribution for the reference system and a system that is parameterized with $(\theta_x, \theta_y, \theta_z)$. The pink cross indicates the parameters of the reference system. The maximum value from left to right 4223, 6463, and 7276. We use equal sized reconstruction and reference time series analog to \cite{HAA15}, which increase the posterior shape in contrast to the paper. Changes in the transition kernel variance $\Sigma_F$ had no major influence. We used the Lorenz63 with an observation time $\Delta t = 0.02$. For all parameters see table \ref{tab:parameteres} first row and table \ref{tab:mcmc}.}
    \label{fig:mcmc_02_original}
\end{figure}

\section{Parameters}
\begin{table}[H]
\begin{tabular}{|l|l|l|l|l|l|l|l|l|l|l|l|}
\cline{1-10}
\begin{tabular}[c]{@{}c@{}}Fig. / Tab.\end{tabular}                                                                                                                                                                                                                                                                                                                                                                                                                                                                                                                                                     & Model                                                                                                    & \begin{tabular}[c]{@{}l@{}}Number\\ of Obs., \\ $N_O$\end{tabular} & \begin{tabular}[c]{@{}l@{}}Integration \\ Method\end{tabular}   & \begin{tabular}[c]{@{}l@{}}Obs. \\ Interval, \\$\Delta t$\end{tabular} & \begin{tabular}[c]{@{}l@{}}Integration \\ invernal, $h$\end{tabular} & \begin{tabular}[c]{@{}l@{}}Nodes, \\ $N$\end{tabular} & \begin{tabular}[c]{@{}l@{}}Input \\ Scaling, $k$\end{tabular} & Regulizer          & \begin{tabular}[c]{@{}l@{}}GCI \\ Subsample- \\rate, $c_\text{sub}$\end{tabular}  \\ \cline{1-10}
 \begin{tabular}[c]{@{}l@{}}
 
 \ref{tab:thresholds_dt_02}, \ref{fig:mcmc_02_original}, \ref{adx:fig:quantiles_dt_02_:Lorenz63}, \\\ref{adx:fig:spectral_radius_dt_02:Lorenz63},
 \ref{adx:fig:spectral_radius_threshold_dt_02:Lorenz63}
 
 \end{tabular} & Lorenz63                                                                                                 & 5000                                                                    & \begin{tabular}[c]{@{}l@{}}scipy.integrate\\.RK45\cite{scipy}\end{tabular} & 0.02                                                                        & 0.001                                                                & 20                                                    & 0.4                                                           & $1 \times 10^{-7}$ & 1/15 \\ \cline{1-10}
 \begin{tabular}[c]{@{}l@{}} 
 
 \ref{tab:thresholds_1},\ref{adx:fig:quantiles_dt_:Lorenz63},
 \ref{adx:fig:spectral_radius_dt:Lorenz63}, \\\ref{adx:fig:spectral_radius_threshold_dt:Lorenz63}
 
 \end{tabular}                                                                                                                                                                                                                                                                                                   & Lorenz63                                                                                                 & 5000                                                                    & \begin{tabular}[c]{@{}l@{}}scipy.integrate\\.RK45\cite{scipy}\end{tabular} & 0.1                                                                         & 0.001                                                                & 20                                                    & 1.0                                                           & $5 \times 10^{-6}$ &  1/3  \\ \cline{1-10}
\begin{tabular}[c]{@{}l@{}}

\ref{tab:thresholds_dt_02}, \ref{adx:fig:sweep_shring_and_inflate_02:FirstOrderODESprottB} - \ref{adx:fig:sweep_shring_and_inflate_02:Roessler76}, \\
\ref{adx:fig:quantiles_dt_02_:FirstOrderODESprottB} - \ref{adx:fig:quantiles_dt_02_:FirstOrderODESprottD}, \ref{adx:fig:quantiles_dt_02_:Roessler76} \\
\ref{adx:fig:spectral_radius_dt_02:FirstOrderODESprottB} - \ref{adx:fig:spectral_radius_dt_02:FirstOrderODESprottD}, \ref{adx:fig:spectral_radius_dt_02:Roessler76} \\
\ref{adx:fig:spectral_radius_threshold_dt_02:FirstOrderODESprottB} - \ref{adx:fig:spectral_radius_threshold_dt_02:FirstOrderODESprottD}, \ref{adx:fig:spectral_radius_threshold_dt_02:Roessler76} \\

\end{tabular}                                                                                                                                                                                                                                                                                                               & \begin{tabular}[c]{@{}l@{}}Sprott B\\ \hline Sprott C\\ \hline Sprott D\\ \hline Roessler76\end{tabular} & 5000                                                                    & \begin{tabular}[c]{@{}l@{}}scipy.integrate\\.RK45\cite{scipy}\end{tabular} & 0.2                                                                         & 0.02                                                                 & 20                                                    & 0.4                                                           & $1 \times 10^{-7}$ &   1/15 \\ \cline{1-10}
\begin{tabular}[c]{@{}l@{}}

\ref{tab:thresholds_1}, \\
\ref{adx:fig:quantiles_dt_:FirstOrderODESprottB} - \ref{adx:fig:quantiles_dt_:FirstOrderODESprottD}, \ref{adx:fig:quantiles_dt_:Roessler76} \\
\ref{adx:fig:spectral_radius_dt:FirstOrderODESprottB} - \ref{adx:fig:spectral_radius_dt:FirstOrderODESprottD}, \ref{adx:fig:spectral_radius_dt:Roessler76} \\
\ref{adx:fig:spectral_radius_threshold_dt:FirstOrderODESprottB} - \ref{adx:fig:spectral_radius_threshold_dt:FirstOrderODESprottD}, \ref{adx:fig:spectral_radius_threshold_dt:Roessler76} \\

\end{tabular}                                                                                                                                                                                                                                                                                                  & \begin{tabular}[c]{@{}l@{}}Sprott B\\ \hline Sprott C\\ \hline Sprott D\\ \hline Roessler76\end{tabular} & 5000                                                                    & \begin{tabular}[c]{@{}l@{}}scipy.integrate\\.RK45\cite{scipy}\end{tabular} & 1.0                                                                         & 0.02                                                                 & 20                                                    & 1.0                                                           & $5 \times 10^{-6}$ & 1/3 \\  \cline{1-10}

\end{tabular}
\caption{Hyperparameters of the Solver, Reservoir, and GCI Measures.}
\label{tab:mcmc}
\end{table}

\begin{table}[H]
\centering
\begin{tabular}{|l|l|l|l|l|l|}
\hline
Figure                                                                      & Model                                                                                                                                     & $\Sigma_F$ & $s_d$       & $t_0$ & $\epsilon$ \\ \hline
\begin{tabular}[c]{@{}l@{}} \ref{fig:mcmc_02_original}\end{tabular} & \begin{tabular}[c]{@{}l@{}} Lorenz63 \end{tabular} & $ 0.01 \times\mathbbm{1}$     & $2.4^2 / 3$ & $50$  & $0$        \\ \hline
\end{tabular}
\caption{Adaptive Metropolis Hastings Parameters}
\label{tab:parameteres}
\end{table}

\section{Further Models}
\label{apd:further_models}

\subsection{Sprott B}

\begin{align*}
\frac{du_x}{dt} &= u_y \cdot u_z \\
\frac{du_y}{dt} &= u_x - u_y \\
\frac{du_z}{dt} &= 1 - u_x \cdot u_y
\end{align*}

\subsection{Sprott C}

\begin{align*}
\frac{du_x}{dt} &= u_y \cdot u_z \\
\frac{du_y}{dt} &= u_x - u_y \\
\frac{du_z}{dt} &= 1 - u_x^2
\end{align*}

\subsection{Sprott D}

\begin{align*}
\frac{du_x}{dt} &= -u_y \\
\frac{du_y}{dt} &= u_x + u_z \\
\frac{du_z}{dt} &= u_x \cdot u_z + 3 \cdot u_y^2
\end{align*}

\subsection{Roessler 76}

\begin{align*}
\frac{du_x}{dt} &= -(u_y + u_z) \\
\frac{du_y}{dt} &= u_x + 0.2 \cdot u_y \\
\frac{du_z}{dt} &= 0.2 + u_z \cdot (u_x - 5.7).
\end{align*}

\section{Validation}

\newcommand{\validationCaption}[2]{For the #1, a one-dimensional for different scaled time series (x-axis) and different initial conditions on the attractor (25, 50, and 75 quantiles). The time series were scaled around the center of mass. Further, we used fixed $(\theta_x, \theta_y, \theta_z) = (10, 28, \frac{8}{3})$, an observation interval of $\Delta t = #2$, a simulation interval of $dt = 0.001$, and created $N=5000$ observations. We train all measures with reference hyperparameter $\theta_x=10$ (black solid line) and the thresholds are determined (dotted black lines). We see that all measures are optimal for the reference system. Using the threshold, the interval where the rejection is less than 75\% (gray area) is defined. It is the largest for the Log GCI, making it the least sensitive measure, and the smallest for the AExc. For all parameters see table \ref{tab:parameteres} third row.}

\foreach \Model/\Fname in {Sprott B/FirstOrderODESprottB, Sprott C/FirstOrderODESprottC, Sprott D/FirstOrderODESprottD, Roessler76/Roessler76}{

    \begin{figure}[H]
    \centering
    \includegraphics[width=0.8\textwidth]{figures/supplemental/sweep_shring_and_inflate_02_\Fname.pdf}

    \caption{\validationCaption{\Model}{0.02}}
    \label{adx:fig:sweep_shring_and_inflate_02:\Fname}
    \end{figure}
}

\section{Sparse Observation Time}

\subsection{Thresholds}

\begin{table}[H]
    \centering
    \begin{tabular}{|l|r|r|r|r|}
    \hline
                &   AExc &   ADev &     TVar &   GCI \\\hline
 Sprott B   &      0 & 231471 & 0.699 & 18.99 \\\hline
 Sprott C   &      0 & 209421 & 0.598 & 21.07 \\\hline
 Sprott D   &      0 & 129971 & 0.804 & 18.66 \\\hline
 Lorenz63   &      0 &  86332 & 0.840 & 18.23 \\\hline
 Roessler76 &      0 & 111894 & 0.880 & 18.68 \\\hline
    \end{tabular}
    \caption{Empirically determined thresholds for different Measures and dynamical systems by repeatably measuring $50000$ reference time series with an observation time of $\Delta t = 0.1$ and $5000$ observations per time series. The threshold is determined as the 0.95-quantile of the resulting distribution. In the paper, all measures except the GCI use this empirical threshold for the hypothesis testing. A theoretical value of $18.307$ is used for the GCI since the distribution is known. By definition, all values of the AExc are 0, and thus, the threshold as well. }
    \label{tab:thresholds_1}
\end{table}

\subsection{Measure Ranges}

\newcommand{\quantileCaption}[2]{Using the #1 model and an observation time of $\Delta t = #2$, the subfigures on the right show histograms of the values the different measures can produce, if the a references trajectory from the true system is evaluated. The vertical dotted/dashed lines indicate the threshold, defined via the quintiles, at which we exude something to be significantly different from the true system. The solid black line is a chi-squared distribution with 10 degrees of freedom and describes the reference distribution of the GCI if the samples are independent. The subfigures on the left provide a detailed view for the acceptance region. For all parameters see table \ref{tab:parameteres} second and fourth row. For the threshold see table \ref{tab:thresholds_1}. }

\foreach \Model/\Fname in {Sprott B/FirstOrderODESprottB, Sprott C/FirstOrderODESprottC, Sprott D/FirstOrderODESprottD, Lorenz63/Lorenz63, Roessler76/Roessler76}{

    \begin{figure}[H]
    \centering
    \includegraphics[width=1.0\textwidth]{figures/supplemental/quantiles_\Fname.pdf}

    \caption{\quantileCaption{\Model}{0.1}}
    \label{adx:fig:quantiles_dt_:\Fname}
    \end{figure}
}

\subsection{Measure vs Spectral Radius}

\newpage

\newcommand{\measureCaption}[2]{In contrast to the paper, this plot uses uses a large time step, and thus, has spare observations. For the parameters see table \ref{tab:parameteres} second and fourth row. The figure shows the value of the Valid Prediction Time (VPT), the percentage of bounded and oscillating solutions, the four measures vs the spectral radius for autonomous trajectories from a 20-node Reservoir for the #1 system. For the four measures only the bounded and oscillating were used. For the number of bounded and oscillating we show the percentage (blue line), while for all other values the 25, 50, and 75 quantiles are plotted (blue line and gray area). For each spectral radius, we used 1000 reservoirs with different input weights and training data.
 For all parameters see table \ref{tab:parameteres} second and fourth row.
}

\foreach \Model/\Fname in {Sprott B/FirstOrderODESprottB, Sprott C/FirstOrderODESprottC, Sprott D/FirstOrderODESprottD, Lorenz63/Lorenz63, Roessler76/Roessler76}{

    \begin{figure}[H]
    \centering
    \includegraphics[width=0.9\textwidth]{figures/supplemental/spectral_radius_dt_\Fname.pdf}

    \caption{\measureCaption{\Model}{0.1}}
    \label{adx:fig:spectral_radius_dt:\Fname}
    \end{figure}
}

\subsection{Rejection Rate vs Spectral Radius}

\newpage

\newcommand{\thresholdCaption}[2]{In contrast to the paper, this plot uses uses a large time step, and thus, has spare observations. For the parameters see table \ref{tab:parameteres} second and fourth row. The figure shows rejection rate of the Valid Prediction Time (VPT), the percentage of bounded and oscillating solutions, the four measures vs the spectral radius for autonomous trajectories from a 20-node Reservoir for the #1 system. For the four measures only the bounded and oscillating were used. For the VPT we show the 25, 50, and 75 quantiles are plotted (blue line and gray area), while for the other values the percentage was used (blue line). The threshold can be found in table \ref{tab:thresholds_1}. For each spectral radius, we used 1000 reservoirs with different input weights and training data. For all parameters see table \ref{tab:parameteres} second and fourth row. For the threshold see table \ref{tab:thresholds_1}.}
\foreach \Model/\Fname in {Sprott B/FirstOrderODESprottB, Sprott C/FirstOrderODESprottC, Sprott D/FirstOrderODESprottD, Lorenz63/Lorenz63, Roessler76/Roessler76}{

    \begin{figure}[H]
    \centering
    \includegraphics[width=0.9\textwidth]{figures/supplemental/spectral_radius_threshold_dt__\Fname.pdf}

    \caption{\thresholdCaption{\Model}{0.1}}
    \label{adx:fig:spectral_radius_threshold_dt:\Fname}
    \end{figure}
}

\section{Paper Observation Time}

\subsection{Thresholds}

\begin{table}[H]
    \centering
    \begin{tabular}{|l|r|r|r|r|}
    \hline
                &   AExc &   ADev &     TVar &   log GCI \\\hline
 Sprott B   &      0 & 231591 & 0.716 & 18.71 \\\hline
 Sprott C   &      0 & 209552 & 0.614 & 18.56 \\\hline
 Sprott D   &      0 & 130243 & 0.838 & 18.43 \\\hline
 Lorenz63   &      0 &  86465 & 0.848 & 18.67 \\\hline
 Roessler76 &      0 & 111976 & 0.883 & 19.11 \\\hline
\end{tabular}
\caption{Empirically determined thresholds for different Measures and dynamical systems by repeatably measuring $50000$ reference time series with an observation time of $\Delta t = 0.02$ and $5000$ observations per time series. The threshold is determined as the 0.95-quantile of the resulting distribution. In the paper, all measures except the GCI use this empirical threshold for the hypothesis testing. A theoretical value of $18.307$ is used for the GCI since the distribution is known. By definition, all values of the AExc are 0, and thus, the threshold as well.}
\label{tab:thresholds_dt_02}
\end{table}

\subsection{Measure Ranges}

\newcommand{\quantileCaptionDense}[2]{Using the #1 model and an observation time of $\Delta t = #2$, the subfigures on the right show histograms of the values the different measures can produce, if the a references trajectory from the true system is evaluated. The vertical dotted/dashed lines indicate the threshold, defined via the quintiles, at which we exude something to be significantly different from the true system. The solid black line is a chi-squared distribution with 10 degrees of freedom and describes the reference distribution of the GCI if the samples are independent. The subfigures on the left provide a detailed view for the acceptance region. For all parameters see table \ref{tab:parameteres} first and third row. For the threshold see table \ref{tab:thresholds_dt_02}. }

\foreach \Model/\Fname in {Sprott B/FirstOrderODESprottB, Sprott C/FirstOrderODESprottC, Sprott D/FirstOrderODESprottD, Lorenz63/Lorenz63, Roessler76/Roessler76}{

    \begin{figure}[H]
    \centering
    \includegraphics[width=0.85\textwidth]{figures/supplemental/quantiles_02_\Fname.pdf}

    \caption{\quantileCaptionDense{\Model}{0.02}}
    \label{adx:fig:quantiles_dt_02_:\Fname}
    \end{figure}
}

\subsection{Measure vs Spectral Radius}

\newcommand{\measureCaptionDense}[2]{This plot uses small time step analogs to the paper for the parameters see table \ref{tab:parameteres} first and third row. The figure shows the Valid Prediction Time (VPT), the percentage of bounded and oscillating solutions, the four measures vs the spectral radius for autonomous trajectories from a 20-node Reservoir for the #1 system. For the four measures only the bounded and oscillating were used. For the number of bounded and oscillating we show the percentage (blue line), while for all other values the 25, 50, and 75 quantiles are plotted (blue line and gray area). For each spectral radius, we used 1000 reservoirs with different input weights and training data.
For all parameters see table \ref{tab:parameteres} first and third row. For the threshold see table \ref{tab:thresholds_dt_02}.
}

\foreach \Model/\Fname in {Sprott B/FirstOrderODESprottB, Sprott C/FirstOrderODESprottC, Sprott D/FirstOrderODESprottD, Lorenz63/Lorenz63, Roessler76/Roessler76}{

    \begin{figure}[H]
    \centering
    \includegraphics[width=0.9\textwidth]{figures/supplemental/spectral_radius_dt_02_\Fname.pdf}

    \caption{\measureCaptionDense{\Model}{0.02}}
    \label{adx:fig:spectral_radius_dt_02:\Fname}
    \end{figure}
}

\subsection{Rejection Rate vs Spectral Radius}

\newpage

\newcommand{\thresholdDenseCaption}[2]{This plot uses small time step analogs to the paper for the parameters see table \ref{tab:parameteres} first and third row. The figure shows rejection rate of the Valid Prediction Time (VPT), the percentage of bounded and oscillating solutions, the four measures vs the spectral radius for autonomous trajectories from a 20-node Reservoir for the #1 system. For the four measures only the bounded and oscillating were used. For the VPT we show the 25, 50, and 75 quantiles are plotted (blue line and gray area), while for the other values the percentage was used (blue line). The threshold can be found in table \ref{tab:thresholds_dt_02}. For each spectral radius, we used 1000 reservoirs with different input weights and training data. For all parameters see table \ref{tab:parameteres} first and third row. For the threshold see table \ref{tab:thresholds_dt_02}.}

\foreach \Model/\Fname in {Sprott B/FirstOrderODESprottB, Sprott C/FirstOrderODESprottC, Sprott D/FirstOrderODESprottD, Lorenz63/Lorenz63, Roessler76/Roessler76}{

    \begin{figure}[H]
    \centering
    \includegraphics[width=0.9\textwidth]{figures/supplemental/spectral_radius_threshold_dt__02_\Fname.pdf}

    \caption{\thresholdDenseCaption{\Model}{0.02}}
    \label{adx:fig:spectral_radius_threshold_dt_02:\Fname}
    \end{figure}
}

\bibliography{agluedgebib_export_2025_05_30, references}